\documentclass[a4paper,fleqn]{cas-dc}

\usepackage{amssymb}
\usepackage{amsthm}
\usepackage{longtable}
\usepackage{amsmath}
\usepackage{longtable}
\usepackage{tikz}
\usepackage{rotating}
\usepackage[utf8]{inputenc}
\usepackage{booktabs,comment,xltabular}
\usepackage{multirow}
\usepackage{xcolor}
\usepackage{color, colortbl}
\usepackage{makecell}
\usepackage{enumitem}
\usepackage[numbers]{natbib}
\usepackage{framed}
\usepackage{tcolorbox}
\tcbuselibrary{breakable}

\let\printorcid\relax

\usepackage[colorinlistoftodos]{todonotes}

\theoremstyle{definition}

 \usepackage{lineno}

\def\tsc#1{\csdef{#1}{\textsc{\lowercase{#1}}\xspace}}
\tsc{WGM}
\tsc{QE}

\colorlet{shadecolor}{gray!20}

\usepackage{float}
\usepackage{xurl}

\begin{document}
\let\WriteBookmarks\relax
\def\floatpagepagefraction{1}
\def\textpagefraction{.001}

\shorttitle{Connecting the Dots in Trustworthy Artificial Intelligence}

\shortauthors{N. Díaz-Rodríguez, J. Del Ser et al.}   

\title [mode = title]{
Connecting the Dots in Trustworthy Artificial Intelligence: From AI
Principles, Ethics, and Key Requirements to Responsible AI Systems and Regulation
}

\author[1]{Natalia Díaz-Rodríguez}\cormark[1]
\author[2,3]{Javier {Del Ser}}\cormark[1]
\author[4]{Mark Coeckelbergh}
\author[5,6,7]{Marcos {López de Prado}}
\author[1]{Enrique Herrera-Viedma}
\author[1]{Francisco Herrera}

\address[1]{Department of Computer Science and Artificial Intelligence, DaSCI Andalusian Institute in Data Science and Computational Intelligence, University of Granada, Granada 18071, Spain}
\address[2]{TECNALIA, Basque Research and Technology Alliance (BRTA), 48160 Derio, Spain}
\address[3]{Department of Communications Engineering, University of the Basque Country (UPV/EHU), 48013 Bilbao, Spain}
\address[4]{Department of Philosophy, University of Vienna, Vienna, 1010, Austria}
\address[5]{School of Engineering, Cornell University, Ithaca, NY, 14850, United States}
\address[6]{ADIA Lab, Al Maryah Island, Abu Dhabi, United Arab Emirates}
\address[7]{Department of Mathematics, Khalifa University of Science and Technology, Abu Dhabi, United Arab Emirates}
\cortext[cor1]{These authors contributed equally. Corresponding authors:
nataliadiaz@ugr.es (N. Díaz-Rodríguez), javier.delser@tecnalia.com (J. Del Ser).}
 
\nonumnote{The views expressed in this article are the authors', and are not necessarily the views of the institutions they are affiliated with.}

\begin{abstract}
Trustworthy Artificial Intelligence (AI) is based on seven technical requirements sustained over three main pillars that should be met throughout the system's entire life cycle: it should be (1) lawful, (2) ethical, and (3) robust, both from a technical and a social perspective. However, attaining truly trustworthy AI concerns a wider vision that comprises the trustworthiness of all processes and actors that are part of the system’s life cycle, and considers previous aspects from different lenses. A more holistic vision contemplates four essential axes: the global principles for ethical use and development of AI-based systems, a philosophical take on AI ethics, a risk-based approach to AI regulation, and the mentioned pillars and requirements. The seven requirements (human agency and oversight; robustness and safety; privacy and data governance; transparency; diversity, non-discrimination and fairness; societal and environmental wellbeing; and accountability) are analyzed from a triple perspective: \textit{What} each requirement for trustworthy AI is, \textit{Why} it is needed, and \textit{How} each requirement can be implemented in practice. On the other hand, a practical approach to implement trustworthy AI systems allows defining the concept of \textit{responsibility} of AI-based systems facing the law, through a given auditing process. Therefore, a responsible AI system is the resulting notion we introduce in this work, and a concept of utmost necessity that can be realized through auditing processes, subject to the challenges posed by the use of regulatory sandboxes. 
{\textcolor{black}{Our multidisciplinary vision of trustworthy AI culminates in a debate on the diverging views published lately about the future of AI. Our reflections in this matter conclude that regulation is a key for reaching a consensus among these views, and that trustworthy and responsible AI systems will be crucial for the present and future of our society.}}

\end{abstract}



\begin{keywords}

\sep Trustworthy AI \sep AI Ethics \sep Responsible AI systems \sep AI Regulation \sep Regulatory Sandbox 

\end{keywords}

\maketitle
\section{Introduction} \label{sec:introduction}  

\begin{figure*}[htbp!]
  \centering
  \includegraphics[width=0.9\linewidth]{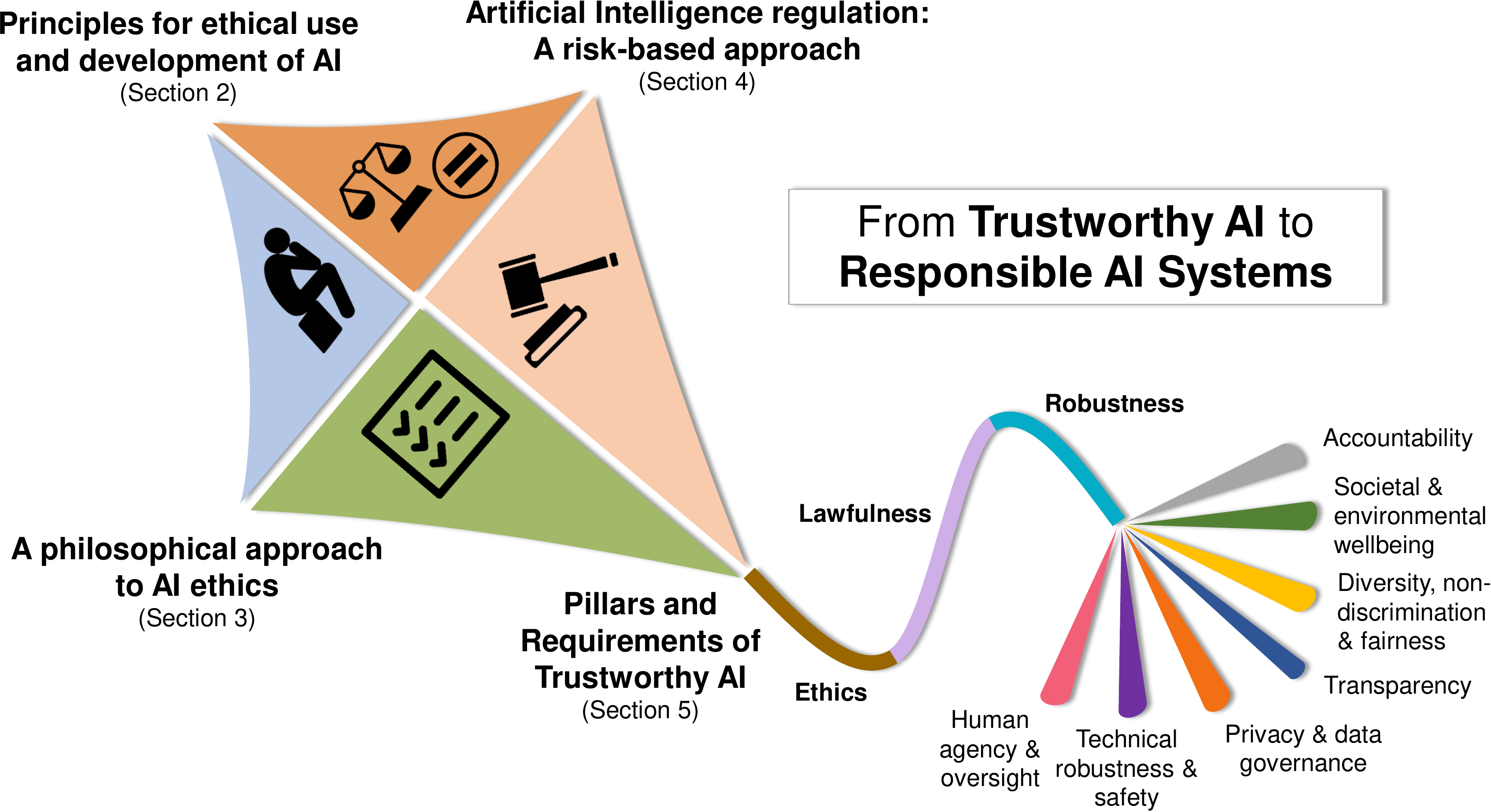}
\caption{Our holistic approach to attain responsible AI systems from trustworthy AI breaks down trustworthy AI into 4 critical axes: assuring the principles for ethical 
development and use of AI, philosophically reflecting on AI ethics, complying with AI regulation, and fulfilling Trustworthy AI requirements.}  \label{fig:cometa}
\end{figure*}

We are witnessing an unprecedented upsurge of Artificial Intelligence (AI) systems. Despite its important historical development, in the last years AI has vigorously entered all professional and social domains of applications, from automation to healthcare, education and beyond. Recently, a family of generative AI (DALL-E\footnote{DALL-E, \url{https://openai.com/product/dall-e-2}, accessed on April 25th, 2023.} \cite{ramesh2021zero}, Imagen\footnote{Google Imagen, \url{https://imagen.research.google/}, accessed on April 25th, 2023.} \cite{saharia2022photorealistic} or large language model products such as ChatGPT\footnote{Chat Generative Pre-Trained Transformer (Chat GPT), \url{https://chat.openai.com/chat}, accessed on April 25th, 2023.}) have sparked a significant amount of debates. These arise as a concern on what this could mean in all fields of application and what impact they could have.

These debates, active for years now, pose questions regarding the ethical aspects and requirements that AI systems must comply with. They emerge from the ethical principles' perspective, from the regulation ones, from what it means to have fair AI, or from the technological point of view, on what an ethical development and use of AI systems really mean. The notion of trustworthy AI has attracted particular interest across the political institutions of the European Union (EU). The EU has intensively worked on elaborating this concept through a set of guidelines based on ethical principles and requirements for trustworthy AI \cite{hleg2019ethics}.

Trustworthy AI is a holistic and systemic approach that acts as prerequisite for people and societies to develop, deploy and use AI systems \cite{hleg2019ethics}. It is composed of three pillars and seven requirements: the legal, ethical, and technical robustness pillars; and the following requirements: human agency and oversight; technical robustness and safety; privacy and data governance; transparency; diversity, non-discrimination and fairness; societal and environmental wellbeing; and accountability. Although the previous definition is based on requirements, there is a larger multidimensional vision. It considers the ethical debate \emph{per se}, the ethical principles and a risk-based approach to regulation, backed up by the EU AI Act \cite{AIA21}.

The goal of this paper is to become a primer for researchers and practitioners interested in a holistic vision of trustworthy AI from 4 axes (Fig. \ref{fig:cometa}): from ethical principles and AI ethics, to legislation and technical requirements. 
According to this vision, our analysis tackles the main aspects of trustworthy AI in a non-exhaustive but technical fashion, by:
\begin{itemize} [leftmargin=*]
    \item \textit{Providing a holistic vision} of the multifaceted notion of trustworthy AI that considers its diverse principles for ethical use and development of AI, 
    seen from international agencies, governments and the industry.  
    \item \textit{Breaking down this multidimensional vision} of trustworthy AI into 4 axes, to reveal the intricacies associated to its pillars, its technical and legal requirements, and what \textit{responsibility} in this context really means. 

    \item \textit{Examining requirements for trustworthy AI}, addressing \textit{what} each requirement actually means, \textit{why} it is necessary and proposed, and \textit{how} it is being addressed technologically. While this paper is not intended to be an exhaustive review, we will delve into an overview of technical possibilities to address the aforementioned seven key requirements for trustworthy AI. 
    
    \item \textit{Analyzing AI regulation from a pragmatic perspective} to understand the essentials of the most advanced legal piece existing so far, the European Commission perspective, and to fully grasp its practical applicability. 

    \item \textit{Defining responsible AI systems} as the result of connecting the many-sided aspects of trustworthy AI above. This is the notion we advocate for, in order to truly attain trustworthy AI. Their design should be guided by regulatory sandboxes.
    
    \item \textit{Dissecting currently hot debates} on the status of AI, the moratorium letter to pause giant AI experiments, the current movements around an international regulation and our positioning based on the previous analyses.
\end{itemize}

By bridging the gap from theory (AI Principles, Ethics, and Key Requirements) to practice (Responsible AI Systems and Regulation), our holistic view offered in this work aims to ultimately highlight the importance of all these elements in the development and integration of human-centered AI-based systems into the everyday life of humans, in a natural and sustainable way.

The paper is organized as follows: Section \ref{sec:ethicalfairAI} revises the most widely recognized AI principles for 
the ethical use and development of AI (axis 1). Section \ref{sec:ethics} considers axis 2: a philosophical approach to AI ethics. Section \ref{sec:regulation} (axis 3) presents the current risk-based viewpoint to AI regulation. Section \ref{sec:tai} analyzes axis 4, i.e., key requirements to implement trustworthy AI from a 
technical point of view. Section \ref{sec:applicability} discusses the practical applicability of trustworthy AI by first closing the loop through the necessary definition of responsible AI systems, and then exposing the requirements for high-risk AI systems to comply with the law. It also proposes the use of regulatory sandboxes as a challenge to test the former requirements via auditing, and a practical blueprint as a case study for AI healthcare. \textcolor{black}{We end this section by elaborating on the needs of emerging AI systems (including general purpose AI and neuroscience technology), which demand evolved evaluation protocols and dynamic regulation. Section \ref{sec:debate} follows by dissecting currently hot debates on the status of AI, from the AI moratorium
letter to regulation as the key for consensus, including a reflection on the gap to be closed between regulation and technological progress}. Finally, Section \ref{sec:concl} draws concluding remarks, and {\color{black} highlights the aforementioned convergence between AI technology and regulation as the beacon for research efforts that safely bring the benefits of this technology to humanity.}

\section{Principles for ethical use and development of Artificial Intelligence
}\label{sec:ethicalfairAI}

A large set of declarations and guidelines for 
the ethical use and development of AI has bloomed. These declarations lead to different similar approaches for introducing sets of principles as a departure point for discussing about the responsible development of AI.  

In this section we will analyze three different principle declarations. 
We will start in Section \ref{sec:unesco} with the general UNESCO Recommendation on the Ethics of Artificial Intelligence, and continue in Section \ref{sec:telefonica} taking a practical perspective from the industry point of view through the principles of Responsible AI by Design in Practice by \emph{Telefónica}, a global telecommunication and media company with large presence in the Spanish and Portuguese markets. Finally, in Section \ref{sec:hleg} we analyze the ethical principles based on fundamental rights associated to the European approach.

\subsection{UNESCO recommendation on the ethics of Artificial Intelligence}
\label{sec:unesco}

In November 2021, UNESCO proposed in Paris a Recommendation on the Ethics of AI. Recognizing that AI can be of great service to humanity and all countries can benefit from it, but also can raise fundamental ethical concerns (can deepen existing divides and inequities in the world), and accounting for the Universal Declaration of Human Rights (1948) and the rest of relevant international recommendations and declarations, the \textit{UNESCO Recommendation on the Ethics of Artificial Intelligence}, 
\cite{unesco2020Recommendation} are the following:
\begin{enumerate}[leftmargin=*]
    \item Proportionality and do no harm: 
    AI methods should not infringe upon the foundational values in these recommendations, should be based on rigorous scientific foundations, and final human determination should apply. 
    \item Safety and security: Unwanted harms such as safety risks, and vulnerabilities to attacks (security risks) should be avoided and addressed throughout the life cycle of AI systems.
    \item Fairness and non-discrimination: AI actors should promote social justice and safeguard fairness. 
    Member States should tackle digital divides ensuring inclusive access and equity, and participation in the development of AI. 
    \item Sustainability: The continuous assessment of the human, social, cultural, economic and environmental impact of AI technology should be carried out with “full cognizance of the implications of AI for sustainability” as a set of constantly evolving goals.
    \item Right to Privacy, and Data Protection: Privacy must be respected, protected and promoted throughout the AI life cycle. 
    \item Human oversight and determination: Member States should ensure that it is always possible to attribute ethical and legal responsibility for any stage of the life cycle of AI systems, as well as in cases of remedy related to AI systems, to physical persons or existing legal entities.
    \item Transparency and explainability: Transparency is necessary for relevant liability regimes to work effectively. 
    AI actors should commit to ensuring that the algorithms developed are explainable, especially in cases that impact the end user in a way that is not temporary, easily reversible or otherwise low risk.
    \item Responsibility and accountability: ``The ethical responsibility and liability for the decisions and auctions based in any way on an AI system should always ultimately be attributable to AI actors'' 
    \item Awareness and literacy: Public awareness and understanding of AI technologies and the value of data should be promoted through open and accessible education, civic engagement, digital skills and AI ethics training. 
    All society should be able to take informed decisions about their use of AI systems and be protected from undue influence.
    \item Multi-stakeholder and adaptive governance and collaboration: ``Participation of different stakeholders throughout the AI system life cycle is necessary for inclusive approaches to AI governance, enabling the benefit to be shared by all, and to contribute to sustainable development''.

\end{enumerate}

The proposed principles are accompanied by values to promote, e.g., human rights and fundamental freedoms. Values and principles are designed to be respected by all actors involved in the AI system life cycle, being amenable of change through amendments to existing and new legislation and business guidelines, since they must comply with international law, the United Nations Charter and Member States.

\subsection{\textit{Telefónica}'s principles of Responsible AI by Design in Practice }
\label{sec:telefonica}

Enterprises also need to cope with and adapt to new demands of AI products and associated risks. The previous recommendations are also aligned with the more generic principles for AI defined by the Berkman Klein Center for Internet \& Society at Harvard University that started being developed since 2016:  \textit{Principled AI} maps \textit{ethical and rights-based approaches to principles for AI} 
to address issues related to the potential threats of AI to both individuals and society as a whole. Derived from these, in industry, e.g., Telefónica defines the so-called 5 principles of \textit{Responsible AI by Design in Practice}  
\cite{benjamins2019responsible} as:
\begin{enumerate}[leftmargin=*]
    \item Fair AI: the output of AI systems must not lead to discrimination.
    \item Transparent and explainable AI: people should know whether they are communicating with a person or an AI-based system.
    \item Human-centered AI (AI for Social Good, Human-centered AI \cite{pisoni2021human}): AI products and services must be aligned with the UN Sustainable Development Goals.
    \item Privacy and security by design: standards should be considered during all life cycles, also from the Responsible Research and Innovation 
    Guidelines \cite{stahl2018ethics}.
    \item Extend them to any third party.
\end{enumerate}

The adoption of these and similar principles is part of new awareness strategies being carried out in companies, sometimes known as \textit{change management}. Telefónica's approach is only one example of such adoption. This implies a change in organizations culture to take into account and implement these principles on a day-to-day basis.

\subsection{Ethical principles based on fundamental rights}
\label{sec:hleg}

In Europe, the foundations of trustworthy AI adhere to the four ethical principles proposed by the European Commission’s High-Level Expert Group (HLEG) \cite{hleg2019ethics}. These are based on fundamental rights, to which AI practitioners should always strive to adhere, in order to ensure the development, deployment and use of AI systems in a trustworthy way. Trustworthy AI is grounded in fundamental rights and reflected by the \textit{European Commission's Ethical Principles}: 
\begin{enumerate}[leftmargin=*]
\item Respect for human autonomy: Ensure freedom and autonomy of humans interacting with AI systems implies humans should keep full and effective self-determination over themselves and the ability to take part on democratic processes; AI systems should not "unjustifiably subordinate, coerce, deceive, manipulate, condition or herd humans, but rather, argument, complement and empower human cognitive, social and cultural skills, leave opportunity for human choice and securing human oversight over work processes" in AI systems, e.g., support humans in the work environment and support the creation of meaningful work.
\item Prevention of harm\footnote{Harm can be individual or collective, can include intangible harm to social, cultural, political or natural environments and all living beings.}: AI systems should not “cause nor exacerbate harm or adversely affect humans”. AI systems should “protect human dignity, mental and physical integrity, be technically robust and assure they are not open to malicious use”. For instance, they should be supervised so they do not exacerbate adverse impacts due to information asymmetries or unequal balance of power. 
\item Fairness: Fairness is closely related to the rights to Non-discrimination, Solidarity and Justice. Although there are many different interpretations of fairness, the European Commission advocates for having both: a) a \textit{substantive} dimension of fairness that "commits to ensure equal and just distribution of benefits and costs, commits to free from unfair bias, discrimination and stigmatization, 
implies respecting the \textit{principle of proportionality between means and ends} and a careful balancing of competing interests and objectives" \cite{hleg2019ethics}. b) a \textit{procedural} dimension allowing to "contest and seek redress against decisions taken by AI systems or who operates them". 
To achieve this, the entity responsible for the decision must be identifiable, while decision making processes should be explainable. 
\item Explainability: Being crucial for building and maintaining users trust in the AI system, explainability means that processes need to be transparent, the capabilities and purpose of AI systems openly communicated, and decision -to the extent possible- explainable to those directly and indirectly affected. 
When a decision cannot be duly contested (e.g., because explaining a particular model output or decision and what combination of input factors contributed to it is not always possible), then other explainability measures may be required (traceability, auditability and transparent communication on the capabilities of the AI system). This will depend on the context and severity of consequences if an output is erroneous. 

\end{enumerate}
These ethical principles are placed in the context of AI systems. They act as ethical imperatives, and advocate for AI systems to strive to improve individual and collective wellbeing. 

As we can see, the mobilization has been worldwide: from the Montréal Declaration for a responsible development of AI -- an initiative of University of Montréal--, to the Ethics of AI recommendations led by international organisations such as UNESCO, passing by the adoption led by industry. All sets of principles share terminology, common grounds on human rights, and agree on the relevance of preserving human decisions and responsibilities, which are the most prominent features of ethics of AI. 

\section{A philosophical approach to Artificial Intelligence ethics}
\label{sec:ethics} 

Ethics is an academic discipline which is a subfield of philosophy and generally deals with questions such as “What is a good action?”, “What is the value of a human life?”, “What is justice?”, or “What is the good life?” \cite{hleg2019ethics}. 

Aligned with the European Commission ethics guidelines \cite{hleg2019ethics}, our ethical vision of AI consists of five main actions \cite{coeckelbergh2020ai}. These can help smooth the way to attain ethical AI. Next, we develop these, taking a philosophical approach to AI ethics: 
\begin{enumerate}[leftmargin=*]
\item  \textit{Use philosophy and science to examine and critically discuss assumptions around the role that AI and humans play in these scenarios and discussions}. For example, one could critically discuss claims that are made about the possibility of Artificial General Intelligence or human-level AI. Large language models, for instance, may give the impression that they have a human-like level of intelligence, but work very differently than the human brain and make many mistakes that humans would not make. This also leads to the question regarding the differences between humans and machine, and is also linked to the question concerning the moral status of AI. For example, it has been claimed that a chatbot was sentient, while it did not meet the criteria for sentience. That being said, it is not always clear what these criteria are. AI makes us re-visit philosophical questions concerning moral status. 
\item 
\textit{Observe 
attentively the nature of AI and which functions it is assigned to perform today within its diversity of applications.} We should recognize the pervasiveness of AI. One reason why it is important to ask ethical questions about AI is that it is pervasive: it is used in many applications such as search, text generation, recommendations for commercial products, and so on. In the ethical analysis, we need to pay attention to the details of each application
    \item \textit{Discuss the most concrete and pressing ethical and social problems that AI presents in terms of how it is being applied today}. AI raises a number of ethical questions such as privacy and data protection, safety, responsibility, and explainability. For example, a chatbot can encourage someone to take their life. Does this mean that the application is unsafe? How can we deal with this risk? And if something happens, who is responsible? Typically, there are many people involved in technological action. It is also important to be answerable to those who are affected by the technology \cite{coeckelbergh2020artificial}, for example in the case of a suicide\textcolor{black}{\footnote{\url{https://coeckelbergh.medium.com/chatbots-can-kill-d82fde5cf6ca}}} the company may need to be answerable to the family of the victim. Furthermore, it is important that when AI offers recommendations for decisions, it is clear on what basis these recommendations and decisions are taken. One problem is that this is usually not clear in the case of deep learning. In addition, there are societal implications such as potential unemployment caused by the automation that is enabled by AI, and the environmental costs of AI and its infrastructures through energy use and carbon emissions linked to the use of the algorithms, the storage of data, and the production of hardware.

\item \textit{Investigate AI policies for the near future}. There are now already many policy documents on AI, for example the Ethics Guidelines for Trustworthy AI produced by the European Commission’s High-Level Expert Group 
on AI  \cite{hleg2019ethics} and the 
Recommendation on the Ethics of Artificial Intelligence \cite{unesco2020Recommendation}. 
These documents need to be critically investigated. For example, in the beginning, less attention was given to environmental consequences of AI. A more general problem is that principles and lists of ethical considerations are not sufficient; there is still a gap between those principles and implementation in the technology, in standards, and in legal regulation. 
\item \textit{Ask ourselves whether the attention that concentrates the public discourse in AI is useful as we face other problems, and whether AI should be our unique focus of attention}. Given that we also face other global problems such as climate change and poverty, it is important to ask the question regarding prioritization: Is AI the most important problem we face? And if not - if, for instance, we insist on climate change being the main and most urgent global problem - how does AI impact and perhaps contribute to this problem, and how can it perhaps help to solve it? Reflection on these challenges will be important in the coming years.
\end{enumerate}

Once expressed the ethics of AI from the philosophical perspective, the next section will analyze the regulation of AI.  

\section{Artificial Intelligence regulation: A risk-based approach} 
\label{sec:regulation} 

In the currently hot debate of AI, a fundamental aspect is regulating AI for it to be righteous. The most advanced regulation to date is the European Commission's AI Act proposal\footnote{On April 27th, 2023, the Members of European Parliament (MEPs) reached a political agreement on the AI Act,  \url{https://www.euractiv.com/section/artificial-intelligence/news/meps-seal-the-deal-on-artificial-intelligence-act/}, accessed on May 1st, 2023.} for the regulation of AI \cite{AIA21}. 

In this section we are presenting AI regulation from two angles; first in Section \ref{sec:regRisk} from the perspective of risk of AI systems and then, in Section \ref{sec:regHRAIS}, we make a deeper analysis into high-risk AI systems.

\subsection{A risk-based approach to regulate the use of Artificial Intelligence systems}\label{sec:regRisk}

The AI Act draft proposal for a Regulation of the European Parliament and of the Council laying down harmonized rules on AI \cite{AIA21} is the first attempt to enact a horizontal AI regulation. The proposed legal framework focuses on the specific use of AI systems. 
The European Commission proposes to establish a technology-neutral definition of AI systems in EU legislation and defines a classification for AI systems with different requirements and obligations tailored to a ``risk-based approach'', where the obligations for an AI system are proportionate to the level of risk that it poses. 

The rules of the AI Act specifically consider the risks created by AI applications by proposing a list of high-risk applications, setting clear requirements for AI systems for high-risk applications, defining specific obligations for AI users and providers of high risk applications, proposing a conformity assessment before the AI system is put into service or placed on the market, proposing enforcement after it is placed in the market, and proposing a governance structure at European and national levels.

The four levels of risk \cite{AIA21} outlined by the AI Act are the following (Figure \ref{fig:AIA}):

\begin{itemize}[leftmargin=*]
    \item \textbf{Minimal or No risk}: The vast majority of AI systems currently used in the EU fall into this category. The proposal allows the free use of minimal-risk AI. Voluntarily, systems providers of those systems may choose to apply the requirements for trustworthy AI and adhere to voluntary codes of conduct (Art. 69 - \textit{Codes of Conduct})\footnote{Codes of conduct are encouraged by the Commission and the Member States to foster the voluntary application to AI systems other than high-risk AI systems (HRAIs) ``on the basis of technical specification and solutions that are appropriate means of ensuring compliance with such requirements in light of the intended purpose of the systems'' (Art. 69).}. When a compliant AI systems presents a risk, the relevant operator will be required to take measures to ensure the system no longer presents that risk, withdraw the system from market, or recall the risk for a reasonable period commensurate with the nature of the risk (Art. 67 - \textit{Compliant AI systems which present a risk}). For instance: AI-enabled video games or spam filters. 
    \item \textbf{Limited risk}: Systems such that users should be aware that they are interacting with a machine so they can take an informed decision to continue or step back. These have to comply with specific information/transparency obligations; for instance, chatbots, 
    and systems generating \textit{deepfakes} or synthetic content.

\begin{figure*}[htb]
  \centering
  \addtocounter{footnote}{-1}
  \includegraphics[width=0.8\linewidth]{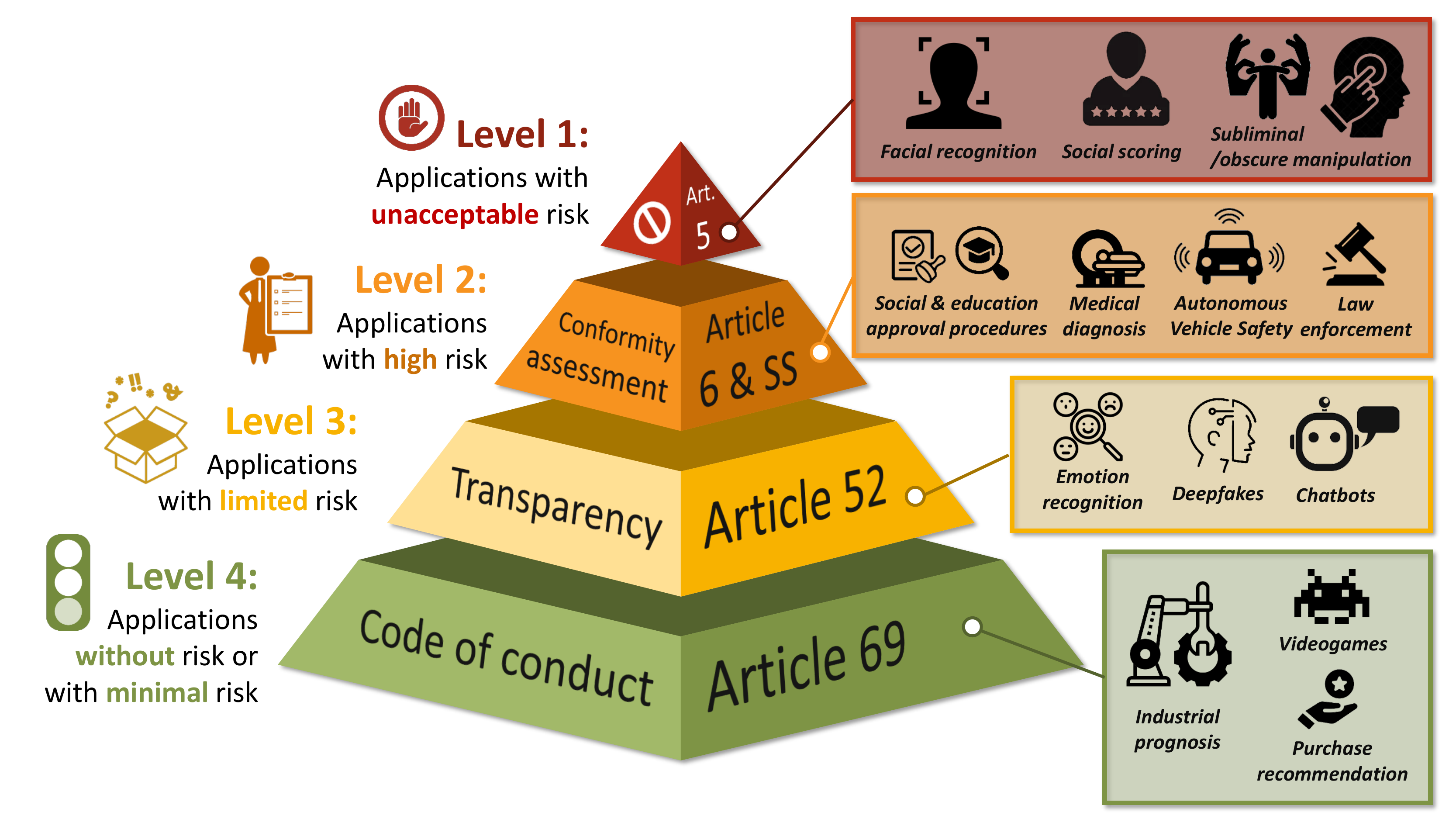}
\caption[Caption for footnotes in caption]%
      {AI Act criticality pyramid and risk-based approach regulatory system for the use of algorithmic systems; SS stands for \emph{subsequent articles} (figure extended from the EU Portal\protect\footnotemark~and inspired from \cite{wahlster2020german} and \cite{edwards2022euAIA}).} 
  \label{fig:AIA}
\end{figure*}

\footnotetext{Regulatory framework proposal on Artificial Intelligence, \url{https://digital-strategy.ec.europa.eu/en/policies/regulatory-framework-ai}, accessed on April 25th, 2023.}

    \item \textbf{High-risk} AI systems (HRAIs): Systems that can have a significant impact on the life chances of a user 
    (Art. 6); they create an adverse impact on people's safety or their fundamental rights\footnote{As protected by the EU Charter of Fundamental Rights (source: \url{https://ec.europa.eu/commission/presscorner/detail/en/QANDA_21_1683}).}. Eight types of systems fall into this category; these are subject to stringent obligations and must undergo conformity assessments before being put on the European market, e.g. systems for law enforcement or access to education. They will always be high-risk when subject to third-party conformity assessment under that sectorial legislation.  
    \item \textbf{Unacceptable risk}: AI systems considered a clear threat to the safety, livelihoods and rights of people will be prohibited in the EU market (Title II — \textit{Prohibited Artificial Intelligence Practices}, Art. 5). For example: Social scoring, 
    facial recognition, dark-patterns and manipulative AI systems, e.g., voice assistance systems that encourage dangerous behavior, or real time remote biometric identification systems in public spaces for law enforcement. 
\end{itemize}

As we can see, very differently to the Chinese, government-centric, approach, or the US industry-owned-data approach to AI, the EU is taking a human-centric approach to regulate the use of AI. This risk scenario-based approach regulates usages rather than models and technology themselves. This is the position we defend.

Since the beginning of 2023, the European Parliament has been considering amendments to the law proposing how to conduct fundamental rights impact assessments and other obligations for users of HRAIs. 
Issues still to be finalized include closing the list of HRAI scenarios above exposed, prohibited practices, 
and details concerning the 
use of copyrighted content to train AI systems\footnote{Financial Times, \emph{European parliament prepares tough measures over use of AI}, \url{https://www.ft.com/content/addb5a77-9ad0-4fea-8ffb-8e2ae250a95a?shareType=nongift}, accessed on April 25th, 2023.} and the regulation of general purpose AI systems \textcolor{black}{(GPAIS)}\footnote{Privacy and Access Council of Canada, \emph{Five considerations to guide the regulation of ``General Purpose AI''}, \url{https://pacc-ccap.ca/five-considerations-to-guide-the-regulation-of-general-purpose-ai/}, accessed on April 25th, 2023.}. The first consideration indicates the EU Parliament will force AI operators to reveal whether they use protected content. The second request emerges from the Privacy and Access Council of Canada, who agrees that GPAIS carry serious risks and harmful unintended consequences, and must not be exempt under the EU AI Act, or equivalent legislation elsewhere. \textcolor{black}{A recent definition of GPAIS can be found in  \cite{campos2023definition}: ``An AI system that can accomplish a range of distinct valuable tasks, including some for which it was not specifically trained''. It has also been referred to as \emph{foundation model} \cite[JRC Glossary, pag. 32]{estevez2022glossary}), but really a GPAIS refers to a model of different nature, beyond the \emph{generative AI} or \emph{foundation models} that can be considered as specific cases of GPAI systems. A point of agreement among all definitions to date is the capability of a GPAIS to accomplish tasks beyond those for which it was originally trained. This is one of the main reasons why GPAIS have become a pivotal topic of debate in what refers to AI regulation. Section \ref{sec:65} will delve further into this.}

\subsection{ 
High-risk Artificial Intelligence systems }\label{sec:regHRAIS}

The European AI Act is predicted to become the global standard for AI regulation\footnote{The EU AI Act’s Risk-Based Approach: High-Risk Systems and What They Mean for Users, \url{https://futurium.ec.europa.eu/en/european-ai-alliance/document/eu-ai-acts-risk-based-approach-high-risk-systems-and-what-they-mean-users}, accessed on April 25th, 2023.} by unifying within a single framework the concept of \emph{risk acceptability} and the \emph{trustworthiness} of AI systems by their users \cite{laux2022trustworthy}. 
The risk-based approach of the AI Act specifically categorizes as HRAIs the following eight kind of AI systems (AI Act, 
\textit{Annex III - High-risk AI systems  referred to in Art. 6(2)} \cite{AIA21}):
\begin{enumerate}[leftmargin=*]
    \item Surveillance systems (e.g., biometric identification and facial recognition systems for law enforcement)
    \item Systems intended for use as security components in the management and operation of critical digital infrastructures (road traffic and water, gas, heat and electricity supply). 
    \item Systems to determine access, admission or assignment of people to educational institutions or programs or to evaluate people (for the purpose of evaluating learning outcomes, learning processes or educational programs). 
    \item Systems intended to be used for recruitment or selection of personnel, screening or filtering of applications and evaluation of candidates, or systems for making decisions on promotion and termination of contractual relationships, assignment of tasks based on individual behavior and the evaluation of performance and behavior.
    \item Systems for assessing the eligibility for public benefits or assistance, assessing creditworthiness or establishing credit scores. Systems for dispatching or prioritizing emergency first response services (firefighters, medical first aid, etc.).
    \item Systems to assess the risk of a person committing crime or recidivism, or the risk that he or she is a potential offender.  
    \begin{itemize}[leftmargin=*]
        \item Systems intended for use as polygraphs or to detect emotional state, or to assess the reliability of evidence in the course of an investigation or prosecution of crime. 
        \item Systems for predicting the occurrence or re-occurrence of crimes based on profiles of people or assessment of personality traits and characteristics or past criminal behavior. 
        \item Systems for profiling individuals by law enforcement authorities in the course of detecting, investigating or prosecuting crimes. 
    \end{itemize}
    \item Systems intended for use by competent public authorities (such as polygraphs or to detect the emotional state of individuals):
    \begin{itemize}[leftmargin=*]
        \item Risk assessment systems, including security risks, irregular immigration or health risk posed by a person seeking to enter a member state. 
        \item Systems for the examination of applications for asylum, visas and residence permits and claims associated with the eligibility of people applying for status. 
    \end{itemize}
    \item Systems intended for the administration of justice and democratic processes (intended to act on behalf of the authorities in the administration of justice for the interpretation of acts or law and the application of the law to a particular set of facts, or evaluation of reliability of evidence).  
\end{enumerate}  

One fact worth noting in the AI Act is its special emphasis on the importance of taking into account, when classifying AI systems, the result of the AI system in relation with the decision or action taken by a human, as well as the immediacy of its effect (AI Act Intro, (32) \cite{AIA21}).

\section{Trustworthy Artificial Intelligence: Pillars and Requirements}
\label{sec:tai} 

In a technical sense, trustworthiness is the confidence of whether a system/model will act as intended when facing a given problem \cite{tjoa2020survey}. This confidence generates trust in the user of the model (the \emph{audience}), which can be supported from multiple perspectives. For instance, trust can be fostered when a system provides detailed explanations of its decisions \cite{doran2017does}. As Lipton puts it, a person can be more confident when using a model if he/she understands how it works and how it produces its decisions \cite{lipton2018mythos}. Likewise, trust can be bolstered if the user is offered guarantees that the model can operate robustly under different circumstances, that it respects privacy, or that it does not get affected by biases present in the data from which it learns.

Trustworthiness is, therefore, a multifaceted requisite for people and societies to develop, deploy and use AI systems, and a \emph{sine qua non} condition for the realization of the \textit{potentially vast social and economic benefits} AI can bring \cite{hleg2019ethics}. Moreover, trustworthy does not concern only the system itself, but also other actors and processes that take their part during the AI life cycle. This requires a holistic and systemic analysis of the pillars and requirements that contribute to the generation of trust in the user of an AI-based system. 

This section addresses this need by first dissecting the three pillars that set the basis for trustworthy AI -- namely, lawfulness, ethics and robustness (Subsection \ref{sec:pillars}) -- followed by a thorough analysis of the seven requirements proposed by the European Commission's High-Level Expert Group (HLEG): human agency and oversight (Subsection \ref{sec:p1}); technical robustness and safety (Subsection \ref{sec:p2}); privacy and data governance (Subsection \ref{sec:p3}); Transparency (Subsection \ref{sec:p4}); diversity, non-discrimination and fairness (Subsection \ref{sec:p5}); societal and environmental wellbeing (Subsection \ref{sec:p6}); and accountability (Subsection \ref{sec:p7}). Definitions (\emph{what does the requirement stand for?}), motivations (\emph{why is the requirement relevant for trustworthiness?}) and a short glimpse at methodologies (\emph{how can the requirement be met in AI-based systems?}) will be given for each of these requirements in their respective sections.

\subsection{The three pillars of trustworthy Artificial Intelligence} \label{sec:pillars}

In general, a pillar can be understood as a fundamental truth of a given idea or concept, from which key requirements to realize the idea can be formulated. Similarly to construction engineering, pillars are essential for building up the concept of trustworthy AI: each pillar is necessary but not sufficient on its own to achieve trustworthy AI. Key requirements can contribute to one or several pillars, just like construction elements such as concrete, formwork or cantilevers are used to help pillars support the structure of the building. These requirements must be continuously ensured throughout the entire life cycle of AI systems, through methodologies that must not only be technical, but also involve human interaction. 

According to the EU Ethical Guidelines for Trustworthy AI \cite{hleg2019ethics}, pillars of trustworthy AI systems are defined as three basic properties that such systems should possess:
\begin{itemize}[leftmargin=*]
\item Pillar 1: \emph{Lawful}. Trustworthy AI systems should comply with applicable laws and regulations, both horizontally (i.e. the European General Data Protection Regulation) and vertically (namely, domain-specific rules that are imposed in certain high-risk application domains, such as medical or finance). 

\item Pillar 2: \emph{Ethical}. Besides their compliance with the law, trustworthy AI systems should also adhere to ethical principles and values. The rapid technological development of current AI-based system rises ethical questions that are not always addressed synchronously by regulatory efforts. The democratized usage of large language models and misinformation using \textit{deepfakes} are two avant-garde exponents of the relevance of Ethics as one of the pillars of trustworthy AI.   

\item Pillar 3: \emph{Robust}. Trustworthy AI systems should guarantee that they will not cause any unintentional harm, working in a safe and reliable fashion from both technical (performance, confidence) and social (usage, context) perspectives. 
\end{itemize}

Trustworthy AI stands on these three pillars. Ideally, they should act in harmony and pushing in synergistic directions towards the realization of trustworthy AI. However, tensions may arise between them: for instance, what is legal is not always ethical. Conversely, ethical issues may require the imposition of law amendments that become in conflict with prevalent regulations. Trustworthy AI must guarantee ethical principles and values, obey the laws, and operate robustly so as to attain its expected impact on the socioeconomic environment in which it is applied. 
\begin{figure*}[htbp!]
  \centering
  \includegraphics[width=0.8\linewidth]{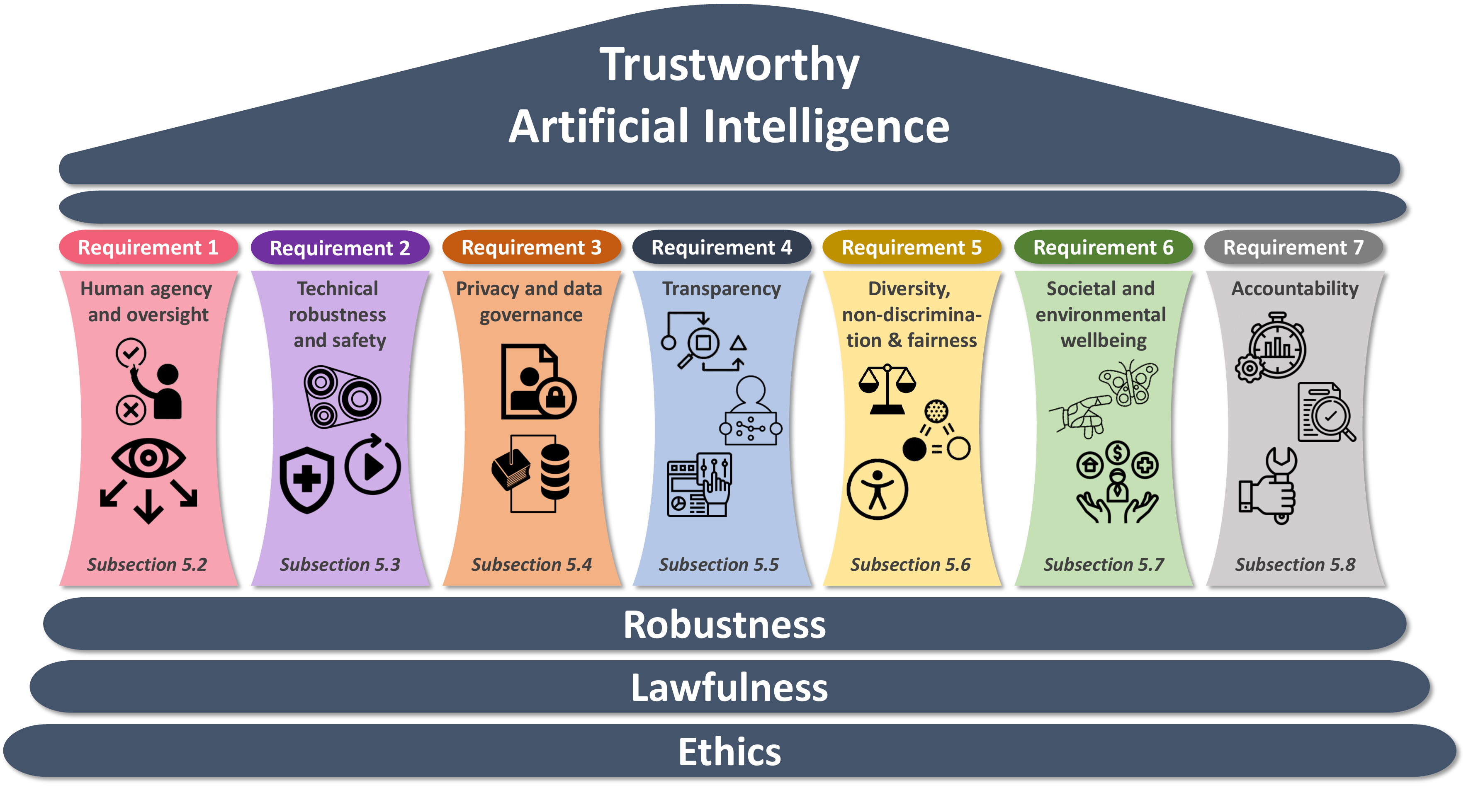}
\caption{Pillars and requirements of Trustworthy AI \cite{hleg2019ethics}.}  \label{fig:pillars_reqs}
\end{figure*}

The above three pillars lie at the heart of the HLEG guidelines \cite{hleg2019ethics}, which establish the seven requirements for trustworthy AI. As depicted in Figure \ref{fig:pillars_reqs}, each requirement spans several components or \emph{dimensions} in which the requirement becomes of special relevance for the design and operation of an AI-based system. Apart from recommending technical and non technical methods, the guidelines also include an Assessment List for Trustworthy Artificial Intelligence, ALTAI, for self-assessment of AI systems and for making the 7 requirements operative \cite{hleg2020altai}. 

The next subsections elaborate in detail on these requirements:

\subsection{Requirement 1: Human agency and oversight} \label{sec:p1}

\vspace{2mm}
\paragraph{WHAT does it mean?} AI systems should empower human beings, allowing them to make informed decisions and fostering their fundamental rights. At the same time, proper oversight mechanisms need to be ensured, which can be achieved through human-in-the-loop, human-on-the-loop, and human-in-command approaches. 
In other words, AI-based systems must support human autonomy and decision making.

\paragraph{WHY is it important for trustworthiness?} This requirement is necessary for autonomy and control. The unfair manipulation, deception, herding and conditioning of AI-based systems can be a threat to the individual autonomy, rights and freedom of their users. Therefore, trustworthy AI systems should provide the means for the user to supervise, evaluate and freely adopt/override a decision made by such systems, avoiding decisions that are automatically made without humans being involved in the process.

\paragraph{HOW can this requirement be met in practice?} Two dimensions underlie this first requirement, namely, human agency (Subsection \ref{sec:agency}) and human oversight (Subsection \ref{sec:oversight}). We now analyze different methodological approaches that can be adopted in these two dimensions:

\subsubsection{Human agency}\label{sec:agency}

Mechanisms for human oversight will depend on the area of application and potential risk. For the preservation of human rights, human-compatible \cite{widmer2022towards}, human-centric approaches \cite{lepri2021ethical,pisoni2023responsible,estevez2022glossary}, AI for social good \cite{tomavsev2020ai,pisoni2021human} and human computation or interactive machine learning \cite{holzinger2016interactive} are computing paradigms aligned with this requirement. However, more structured toolkits (along the lines of \cite{wef2019empowering} or C-Suite \cite{wef2022empowering}) will need to be materialized for a smooth domain-specific consideration of this requirement. In terms of technical tools to reach different audiences, language appears as the universal means of communication among humans and machines, and thus, AI models using natural language processing and/or counterfactual and natural language explanations \cite{cambria2023survey} will be relevant to help humans supervise and take the most appropriate decision based on the output of 
AI systems.

\subsubsection{Human oversight}\label{sec:oversight}

Different degrees of human involvement in the supervision of AI-based systems can be specified \cite{hleg2019ethics}: 
\begin{itemize}[leftmargin=*]
\item \emph{Human-in-the-loop} (HITL), which refers to the ability of the supervisor to intervene in every decision cycle of the system being monitored \cite{holzinger2016interactive}.
\item \emph{Human-on-the-loop} (HOTL), which stands for human intervention during the design and monitoring cycles of the AI-based system.
\item \emph{Human-in-command} (HIC), namely, the capability of the supervisor to oversee the overall activity of the AI system including its broader economic, societal, legal and ethical impacts, and ensuring that decisions produced by the AI system can be overridden by the human.
\end{itemize}

Depending on the application under consideration, mechanisms supporting one of the above levels of human oversight can be designed. Methods proposed so far are largely domain-specific, since user-algorithm interfaces vary depending on the capabilities and background of the supervisor and the design of the AI-based solution.

\subsection{Requirement 2: Technical robustness and safety}  
\label{sec:p2}

\vspace{2mm}
\paragraph{WHAT does it mean?} This second requirement includes several functionalities all aligned with the \emph{prevention of unintentional harm} and the \emph{minimization of the consequences of intentional harm}. These include the resilience of AI-based systems (to attacks and security), ensuring fallback plans (in case something goes wrong), general safety, and being accurate, reliable and reproducible. Robustness and safety refer to the need of AI systems to be secure, reliable and robust enough to errors and/or inconsistencies in all phases of the life cycle \cite{floridi2019establishing}.

\paragraph{WHY is it important for trustworthiness?} AI-based systems deployed on real-world scenarios can undergo changes in their operating environment that can induce changes at their inputs (e.g. concept drift). Likewise, such changes can be the result of the interaction of malicious users with the AI-based system in an adversarial fashion. Disregarding whether such changes are intentional or not, the trustworthiness of the AI-based system is subject to the capability of the model to mitigate the impact of these changes in their issued predictions. Likewise, in risk-critical applications trustworthy AI systems should evaluate relevant safety measures and endowed with functionalities to fall back when the AI-based system deviates from its expected behavior as per the monitored measures. Finally, reliability and reproducibility connects tightly with trustworthiness in what refers to the verification of the expected operation and performance of AI-based systems. When AI-based systems are to be used in different contexts and deployed in different systems, such components are vital to ensure that the system at hand resiliently accommodates the differences and particularities that may arise in each context/system, ultimately working as expected.

\paragraph{HOW can this requirement be met in practice?} Methodologies that can be explored to support this requirement can be analyzed over three dimensions: technical robustness (Subsection \ref{sec:rob}), safety (Subsection \ref{sec:safe}) and reproducibility (Subsection \ref{sec:repr}). 

\subsubsection{Technical robustness} \label{sec:rob}

When dealing with an AI-based system, robustness and reliability are properties that refer to the ability of the system to have comparable performance on atypical data with respect to typical operational regimes \cite{mariani2023trustworthy}. Robustness can be established in the face of different circumstances: when we hope a model to be robust, it is due to the fact that the model may degrade, be perturbed or affected during its future usage. It is desirable to have a model that is robust in terms of its generalization or generative capabilities, against adversarial attacks or models, or against data perturbations. 

Systematically, several levels of robustness can be distinguished in AI-based systems \cite{chen2023ai}:
\begin{itemize}[leftmargin=*]
\item Level 0 (\emph{no robustness} or \emph{standard training}): this first level of robustness refers to the one provided by the AI-based system by itself, without any risk mitigation functionalities or additions added to its design. This level concerns generalization capabilities such as being robust to distributional changes caused by spurious features or data instances. Despite the lack of specific risk mitigation measures, some naive information provided by certain naive AI models (e.g. quantification of epistemic confidence) can be exploited to detect when the AI-based system is not working in its expected operational regime.

\item Level 1 (\emph{generalization under distribution shifts}): this second level of robustness considers techniques aimed to mitigate different types of changes in data. Data changes or \emph{shifts} include covariate shift, prior probability shift, concept drift and confounding shift, depending on the change happening in the distribution of the input variables, the output of the model, the statistical relationship between the inputs and outputs, or the change of a variable that influences both inputs and outputs, respectively \cite{varshney2019trustworthy}. In this level we can also place the generalized framework of Out-of-Distribution (OoD) detection \cite{yang2021generalized}, which refers to anomaly detection, novelty detection and open set recognition, the latter referring to the capability of the model to detect, characterize and incorporate new unknown patterns to its knowledge base (e.g. new classes in a classification problem). Level 1 of robustness against these data shifts can be approached by concept drift detection and adaptation techniques, OoD detection methods or class-incremental learning schemes, to mention a few. 

\item Level 2 (\emph{robustness against a single risk}): this third worst-case robustness tackles a single point of risk, e.g., the presence of adversarial examples. Assessing this level requires model inspection and intervention (e.g., active model scanning, probing to find failure cases, adversarial defenses against different attack modes).

\item Level 3 (\emph{robustness against multiple risks}): It extends the former to multiple risks (e.g., common data corruptions, spurious correlations).

\item Level 4 (\emph{universal robustness}): this level is reached by AI-based systems that are proven to be effectively robust to all known risks.

\item Level 5 (\emph{human-aligned and augmented robustness}): it furthers complements level 4 by aligning human-centered values and user feedback, automatically augmenting existing robustness demands as per the requirements, context and usage of the AI-based system. This level should be targeted by high-risk AI-powered applications. 
\end{itemize}

The robustness of the AI-system system should be a core part of the risk management strategy adopted by the owner of the system itself, hence becoming a core part of their accountability procedures. Indeed, AI maintenance frameworks should ease achieving robustness and AI status tracking and control through the AI life cycle \cite{ruospo2023survey}. Monitoring can be produced either passively (by periodically measuring different quantitative metrics related to robustness over the data, model, or both) or actively (emulating the circumstances under which the robustness of the model can be thought to be compromised (e.g. emulated adversarial attack instances or perturbations of known samples). In both cases, AI maintenance frameworks can detect model degradation through time by detecting systematic deviations of the aforementioned metrics in data and models \cite{speakman2023detecting}. Interestingly, areas currently under study in AI research aim in this direction, endowing AI-based systems with the ability to learn continually from infinite streams of varying data \cite{lesort2020continual}, to quantify and communicate their confidence in their outputs \cite{abdar2021review}, or to characterize and consolidate new patterns arising from data over time \cite{parmar2023open}.

We end the discussion about how technical robustness can be supported in AI-based systems by highlighting the potential that techniques used to address other requirements can bring to technical robustness. For instance, explainability techniques can help make models more robust, since they can show which features are more stable to out of distribution changes in the input or adversarial attacks. Likewise, the intensity of changes needed to reach a target adversarial confidence score in counterfactual generation can be a reliable estimator of the extent to which a certain data instance can be considered to be out of distribution \cite{zimmermann2022increasing}. All in all, these examples are a few among the multiple cases in which a functionality added to an AI-based system can simultaneously contribute to several requirements for trustworthiness.

\subsubsection{Safety}\label{sec:safe}

Evolving from a generic Information Technologies context, safety in AI \cite{amodei2016concrete, hendrycks2021unsolved, mohseni2022taxonomy} is developing in relation to the alignment with human values. In this sense, concrete protocols and procedures are challenging to define, but necessary for AI safety. Safety in AI concerns several unsolved research issues \cite{hendrycks2021unsolved}, including:
\begin{itemize}[leftmargin=*]
    \item Attaining robustness as the objective of withstanding hazards, and building systems less vulnerable to adversarial threats such as adversarial perturbations which cause high confidence mistakes, and robust to long tails.
    \item Facilitating tools to inspect AI-based systems, identify hazards and anomalies, calibrate them, identify honest outputs, and detect emergent capabilities. One risk of AI systems that links with the need for safety tools is that they may carry backdoors \cite{gu2019badnets}: backdoored models behave correctly in nearly all scenarios, except in chosen scenarios taught to behave incorrectly due to the training on poisoned data as a way to have backdoors injected. These are problematic, specially in foundational models that serve as the architectural backbone of downstream models, all evolved from originally poisoned data from massive training datasets \cite{hendrycks2021unsolved}.
    \item Defining safety objectives in order to steer models, either internally (how models should learn to guarantee compliance with safety metrics) and externally (how such safety compliance can be safely pursued). Problems in this regard include:    
    \begin{itemize}[leftmargin=*]
    \item Value learning, as the inability of AI systems to code human values (e.g., happiness, sustainability, meaningful experiences or safe outcomes). Although giving open-world inputs to models can partially tell apart pleasant and unpleasant states, utility values of such states are no ground truth values, and are a result of the model's own learned utility function \cite{hendrycks2021aligning}.
    \item Proxy gaming: This is a phenomenon due to the fact that optimizers and adversaries can manipulate objective proxies. As Goodhart's law states, \emph{a measure ceases to be a reliable indicator when it becomes the target}. For instance, proxy gaming occurs as \emph{reward hacking} in reinforcement learning. Similarly, objective countable metrics end up substituting human values when opaque AI models are forced to learn by optimizing a single quantitative measure\footnote{These are also known as \textit{weapons of math destruction} \cite{o2017weapons} that may contain pernicious feedback loops that perpetuate stereotypes and biases \cite{parikh2019addressing} if they do not consider context nor a concrete person's features, but rather those of its neighbors.}. Therefore, merely acquiring a proxy for human values is insufficient: models must also be resilient to solvers seeking to manipulate it. 
    \end{itemize}
\end{itemize}

\subsubsection{Reproducibility} \label{sec:repr}

Once robustness and safety have been addressed, an important dimension in this key requirement for trustworthy AI is reproducibility. It can be defined as the ability of AI experiments to exhibit the same behavior when repeated under the same conditions. Reproducibility is related to \emph{replicability}, which refers to the capability to independently achieve non-identical conclusions that are at least similar while differences in sampling, research procedures and data analyses may exist \cite{estevez2022glossary}. Since both concepts are essential parts of the scientific method, the National Information Standards Organization (NISO) and the Association for Computing Machinery (ACM) redefine these concepts as:
\begin{itemize}[leftmargin=*]
    \item \emph{Repeatability} (same team, same experimental setup), which means that an individual or a team of individuals can reliably repeat his/her/their own experiment.
    \item \emph{Replicability} (different team, same experimental setup): an independent group of individuals can obtain the same result using artifacts which they independently develop in their entirety.
    \item \emph{Reproducibility} (different team, different experimental setup with stated precision): a different independent group can obtain the same result using their own artifacts.
\end{itemize}

It should be clear that when formulated in the context of trustworthy AI systems, one should regard an \emph{experiment} in the above definitions as the performance, robustness and safety evaluation of a given AI-based system. This evaluation can be done by different groups (as in research) or by a certification lab (as in commercial software-based solutions). The extent to which reproducibility can be guaranteed in trustworthy AI systems depends on the confidentiality of the system or the singularity of the experimental setup for which the AI-based system was developed, among other constraining circumstances. For instance, in mild contexts (as in research), reproducibility of experiments by third parties is often favored by public releases of the source code implementing the AI-based system being proposed. 

\subsection{Requirement 3: Privacy and data governance} \label{sec:p3}

\vspace{2mm}
\paragraph{WHAT does it mean?} This requirements assures the respect for privacy and data protection thorough the AI system life cyle (design, training, testing, deployment and operation), adequate data governance mechanisms taking into account the quality and integrity of the data and its relevance to the domain, and also ensures legitimized access to data and processing protocols.

\paragraph{WHY is it important for trustworthiness?} AI systems based on digital records of human behavior can be capable of inferring individual preferences and reveal personal sensitive information such as the sexual orientation, age, gender, religious or political views. Since AI-based systems learn from data, systems must guarantee that such personal information is not revealed while data is processed, stored and retrieved throughout the AI life cycle, facilitating means to trace how data is used (governance) and verifying that protected information is not accessed (privacy awareness) during the life cycle phases. If such guarantees are not provided, AI-based systems will not be trusted by end users, nor will they conform to existing legislation (e.g. the European GDPR). Citizens should have full control over their data, and their data will not be unlawfully or unfairly used to harm or discriminate against them \cite{floridi2019establishing}. This requirement is important to preserve human rights such as the right to privacy, intimacy, dignity or the right to be forgotten. Keeping the usage and scope of the data limited, protected and informed is paramount, since digital information can be used towards clustering a person into profiles that may not reflect reality, while reinforcing stereotypes, historical differences among minorities, or perpetuate historical or cultural biases \cite{o2017weapons}.

\paragraph{HOW can this requirement be met in practice?} In the following we analyze technologies that can maintain data privacy in AI-based systems (Subsection \ref{sec:priv}), and strategies to deal with data governance as quality and integrity processes (Subsection \ref{sec:gov}).

\subsubsection{Data privacy}\label{sec:priv}

In order to land down the data privacy requirement into actual technologies, we emphasize the relevance of Federated learning (FL) \cite{bonawitz2019towards,rodriguez2020federated}, homomorphic computing \cite{marcolla2022survey} and differential privacy (DP) \cite{abadi2016deep} as examples of privacy-aware technologies in the current AI landscape:
\begin{itemize}[leftmargin=*]
\item In FL, a model is trained across multiple decentralized devices without moving the data to a central location. In doing so, instead of delivering all the data to a central server, devices learn models locally using their own data, so that only numerical model updates are sent to the central server. The central server aggregates the updated model parameters from all the devices or servers to create a new model. This allows learning a global model leveraging all data in situations where the data is sensitive. Besides preserving the privacy of local data, FL can reduce communication costs and accelerate the model training. 

\item In homomorphic computing, data can be processed in encrypted form without the need for deciphering it first. As a result, data remains secure and private by performing operations directly on encrypted data. By using specially devised mathematical operations, the underlying structure of data is preserved while it is processed, so that the result of the computation, which is also encrypted, stays the same. Only authorized parties having the decryption key can access this information. Homomorphic computing can be an effective way to implement privacy-aware preprocessing, training and inference in AI-based systems.

\item Finally, DP enables processing and learning from data while minimizing the risk of identifying individuals in the dataset at hand. To this end, DP injects random noise to the data before it is processed. This noise is calibrated to guarantee that the data remains statistically accurate, while concealing any information that could be used to identify individuals and thereby, compromise their privacy. The amount of noise added to data balances between the level of privacy protection provided by DP and the performance degradation of the AI-based system when compared to the case when no noise is injected.
\end{itemize}

By resorting to any of the above technologies (also combinations of them), the privacy of individuals in the datasets is preserved, minimizing their risk of harm.

\subsubsection{Data governance: Quality and integrity of data and access to data}\label{sec:gov}

Data protocols must govern data integrity and access for all individuals even if these are not users of the AI system. Only duly qualified staff, with explicit need and competence, should be allowed to access individuals' data. As a part of AI governance, data governance calls for a broader level regulation than a single country or continent regulation. This context has motivated guidelines and recommendations for AI governance over the years emphasizing on the importance of ensuring data quality, integrity and access. An example can be found in the Universal Guidelines for AI published in 2018 \cite{universalguidelines}, which were endorsed by 70 organizations and 300 experts across 40 countries. In these guidelines, \emph{Data Quality Obligation} was established as one of the principles that should be incorporated into ethical standards, adopted in regulations and international agreements, and embedded into the design of AI-based systems. These recommendations helped inform the OECD AI Principles (2019), the UNESCO Recommendation on AI Ethics (2021), the OSTP AI Bill of Rights (2022), and the EU AI Act and the Council of Europe Convention on AI.

In terms of guidelines to implement data governance, the Information Commissioner's Officer (ICO) has proposed recommendations on \textit{how to use AI and personal data appropriately and lawfully} \cite{ICO}. Among these, there are actions such as taking a risk-based approach when developing and deploying AI -- ``addressing risk of bias and discrimination at an early stage'', ``ensuring that human reviews of decisions made by AI is meaningful'', ``collect only data needed and no more'', and ``working with external suppliers to ensure the use of AI will be appropriate''.

At the European level, the \textit{European Strategy for Data} established in 2020 aims at making the EU a role model for a society empowered by data. This strategy has given rise to the European \textit{Data Governance Act} \cite{datagovernanceACT22} 
to facilitate data sharing across sectors and Member States. In particular, the EU Data Governance Act intends to make public sector data available for re-use, promote data sharing among businesses, 
allow the use of personal data through a ``personal data-sharing intermediary'', help exercising rights under the General Data Protection Regulation (GDPR), and allowing data use on altruistic grounds \cite{datagovernanceACT22}. 

Later in 2022, the European Union strategy for data proposed the \textit{Data Act} \cite{dataACT22}\footnote{Data Act \& Data Act Factsheet, \url{https://digital-strategy.ec.europa.eu/en/policies/data-act}, accessed on April 25th, 2023.}, a regulation harmonizing rules on fair access to and use of data. In practice, this regulation complements the Data Governance Act by specifying who can create value from data and under which circumstances. In practice, the Data Act will take action to 1) increase legal certainty for companies and consumers who generate data, on who can use what data and under which conditions, 2) prevent abuse of contractual imbalances that hinder fair data sharing. 3) provide means to the public sector to access data of interest held by the private sector; 4) set the framework conditions for customers. Therefore, the benefits of the Data Act for consumers and business include, from achieving cheaper aftermarket prices for connected objects, to new opportunities to use services based on data access, and better access to data produced by devices. Serving these two EU regulations, ten European common data spaces exist, ranging from industry to mobility .

\subsection{Requirement 4: Transparency 
} \label{sec:p4}

\vspace{2mm}
\paragraph{WHAT does it mean?} Transparency is the property that ensures appropriate information reaches the relevant stakeholders \cite{mariani2023trustworthy}. When it comes to AI-based systems, different levels of transparency can be distinguished \cite{arrieta2020explainable}: simulatability (of the model by a human), its decomposability (the ability to explain the model behavior and its parts), and algorithmic transparency (understanding the process of the model and how it will act for any output). Another classification establishes transparency at the algorithmic, interaction and social levels \cite{haresamudram2023three}, emphasizing the role of the stakeholder audience to which the explanation is targeted: developer, designer, owner, user, regulator or society. 

\paragraph{WHY is it important for trustworthiness?} In the context of trustworthy AI systems, data, the system itself and AI business models should be transparent. Humans must be informed of systems capabilities and limitations and always be aware that they are interacting with AI systems \cite{hleg2019ethics}. Therefore, explanations should be timely, adapted and communicated to the stakeholder audience concerned (layperson regulator, researcher or other stakeholder), and traceability of AI systems should be ensured. 

\paragraph{HOW can this requirement be met in practice?} The dimensions to be treated within this requirement concern traceability, explainability and communication, which are essential for realizing transparent AI-based systems. In the following we will first explain what traceability stands for (Subsection \ref{sec:trace}), the current state of the art on explainable AI (Subsection \ref{sec:expl}), and mechanisms for communicating AI systems decisions (Subsection \ref{sec:comm}).

 \subsubsection{Traceability}\label{sec:trace}

Traceability is defined as the set of mechanisms and procedures aimed to keep track of the system's data, development and deployment processes, normally through documented recorded identification \cite{estevez2022glossary}. Traceability and logging from the early design stages of the AI-based systems can help auditing and achieving the required level of transparency according to the needs of the concerned audience. 

In this regard, special attention must be paid to \emph{provenance tools} \cite{perez2018systematic}, which ease the traceability or lineage of data and model decisions, hence contributing to the requirement of transparency. In this area, the use of Blockchain mechanisms are promising towards guaranteeing the integrity of data used to train (and explain) machine learning models, i.e., the provenance of databases, their associated quality, bias and fairness. 

 \subsubsection{Explainability}\label{sec:expl}

The so-called eXplainable AI (XAI) \cite{arrieta2020explainable} field is widely and globally being recognized as a crucial feature for the practical deployment of trustworthy AI models. Existing literature and contributions already made in this field include broad insights into what is yet to be achieved \cite{arrieta2020explainable,holzinger2021information,ALI2023101805}. Efforts have been invested towards defining explainability in machine learning, extending previous conceptual propositions and requirements for responsible AI by focusing on the role of the particular audience for which explanations are to be generated \cite{arrieta2020explainable}: \textit{Given an audience, an explainable AI is one that produces details or reasons to make its functioning clear or easy to understand}.

Explainability techniques are blooming as tools to support algorithmic auditing. They have emerged as a necessary step to validate and understand the knowledge captured by black-box models, i.e., a system in which only inputs and outputs are observed without knowing the internal details of how it works. This can be problematic, as we cannot predict how the system may behave in unexpected situations (connecting with the \emph{technical robustness} requirement, Subsection \ref{sec:p2}), or how it can be corrected if something goes wrong (linked to the \emph{accountability} requirement, Subsection \ref{sec:p7}). Explaining which input factors contribute to the decisions of complex black-box algorithms can provide a useful global view of how the model works, jointly with traceability methods and a clear and adapted communication of information to the target audience.

Since the quality of explanations depends on the audience and the motivation for which they are generated, several taxonomies of XAI techniques have been proposed over the years \cite{arrieta2020explainable}. A primary distinction can be done between model-agnostic and model-specific approaches to explaining machine learning models, the difference being whether the XAI technique can be applied to any machine learning model disregarding their structure and learning algorithm. Another distinction can be done between {\color{black}ex-ante and post-hoc XAI techniques, depending on the moment at which explainability is addressed (before or after the model is designed and trained). On one hand, ex-ante techniques (also referred to as the \emph{explainable-by-design} paradigm) make AI models aspire to provide an explanation that avoids the construction of additional models or extra complexity (layers or mechanisms not originally part of the original one), so that explanations are as faithful to the real reasoning carried out by the model as possible. On the other hand, post-hoc} XAI techniques usually add artifacts around the original AI model or build a surrogate of it -- a local approximation or simpler version of the original one -- in order to more easily explain the original one (for example, LIME \cite{ribeiro2016should}). Likewise, some XAI techniques may use external knowledge (e.g. from the web, Wikipedia, forums) \cite{rajani2019explain}, for instance, to explain language models or dialogue models that interactively answer questions about a model’s particular decision. 

Other criteria to categorize XAI tools can be formulated, such as the format of the issued explanations (e.g., attribution methods \cite{abhishek2022attribution}, counterfactual studies \cite{guidotti2019factual}, simplified model surrogates \cite{van2021evaluating}) or the hybridization of explanations expressed in different modalities, such as visual and textual (e.g., linguistic summaries \cite{kaczmarek2022plenary}, ontologies \cite{bourgeais2022graphgonet}, or logical rules defined on top of knowledge graphs \cite{diaz2022explainable}, to cite a few). Natural language explanations \cite{salewski2022clevr, cambria2023survey}, {\color{black}quantitative measures of the quality of explanations \cite{vilone2021notions,sevillano23}}, and models that support their learning process with formal symbolic basis such as language, rules, compositional relationships or knowledge graphs (neural-symbolic learning and reasoning  \cite{diaz2022explainable}) are key for explanations to be understood by non-expert audience. These interfaces allow such users to assess the operation of the model in a more intelligible fashion, hence supporting the \emph{human agency and oversight} requirement for trustworthy AI systems (Subsection \ref{sec:p1}). 

\subsubsection{Communication}\label{sec:comm}

The third dimension of transparency is how the audience is informed about the AI-based system, namely, how explanations or information tracked about the system's operation is \emph{communicated} to the user. Humans should know when they are interacting with AI systems, as well as be notified about their performance, instructed about their capabilities, and warned about their limitations. The same holds when conveying the model's output explanation and its functioning to the user. The adaptation of the explanation must be in accordance to the specifics of the AI system being explained and the cognitive capabilities (knowledge, background expertise) of the audience. 

Therefore, \emph{communication} is a crucial dimension, so that all aspects related to transparency are delivered to the audience in a form and format adapted to their background and knowledge. This is key to attain trust in the audience about the AI-based system at hand.

\subsection{Requirement 5: Diversity, non-discrimination and fairness}
 \label{sec:p5}
\vspace{2mm}
\paragraph{WHAT does it mean?}  
This requirement contemplates different dimensions: the  
avoidance of unfair bias, diversity fostering, accessibility to all regardless any disability, and the involvement of stakeholders in the entire AI system life cycle. All these dimensions of this manifold requirement share an ultimate purpose: to ensure that AI-based systems do not deceive humans nor limit their freedom of choice without reason. Therefore, it is a requirement tightly linked to the ethical and fairness principles that underlie trustworthiness in AI (Section \ref{sec:ethicalfairAI}).

\paragraph{WHY is it important for trustworthiness?} This requirement is necessary to broaden the impact of AI to all social substrates, as well as to minimize the negative effects that automated decisions may have in practice if data inherits biases hidden in the data from which models are learned. Unfair bias in data must be avoided as decisions drawn by a model learned from such data could have multiple negative implications, including the marginalization of vulnerable groups, the exacerbation of prejudice or discrimination \cite{hleg2019ethics}. 

\paragraph{HOW can this requirement be met in practice?} Methods to tackle this requirement can be classified depending on the specific dimension they support: as such, methods to enforce diversity, non-discrimination, accessibility, universal design and stakeholder participation are briefly revisited in Subsection \ref{sec:div}, whereas Subsection \ref{sec:fairness} describes how to achieve fairness in AI-based systems. Finally, Section \ref{sec:bias} examines mechanisms to avoid unfair bias.

\subsubsection{Diversity, non-discrimination, accessibility, universal design and stakeholder participation}\label{sec:div}

AI systems should take into account all human abilities, skills and requirements, and ensure accessibility to them. Developing methodologies based on the requirement of non-discrimination and bias mitigation is paramount to ensure the alignment of requirements to the compliance with ethical values. Assuring properties of non-discrimination, fairness and bias mitigation restrict the systematic differences treating certain groups (of people or objects) with respect to others \cite{mariani2023trustworthy}. A practical example of recommendation encourages, e.g., hiring from diverse backgrounds, cultures and disciplines to assure opinion diversity.

This requirement involves the inclusion of diverse data and people, and ensures that individuals at risk of exclusion have equal access to AI benefits. Concrete implementations of this requirement range from quantifying the impact of demographic imbalance \cite{hupont2019demogpairs} and the effects of missing data (which, as a matter of fact, has been shown to be beneficial in terms of fairness metrics \cite{fernando2021missing}).

In what refers to \textit{diversity}, it advocates for the needs for heterogeneous and randomly sampling procedures for data acquisition, diverse representation of a population that includes minorities, and the assurance for non-discriminating automated processes that lead to unfairness or biased models. Diversity can be applied at the technical level during model training by penalizing the lack of diverse prototypes on latent space areas with challenging separation between classes \cite{gee2019explaining}. Alternatively, the lack of diversity can be counteracted by means of imbalanced learning or by informing data augmentation. When placing the focus on the solutions of an AI-based system, their diversity is a very relevant component to guarantee non-biased outcomes. Frameworks unifying quality and diversity optimization can guarantee the diversity of generated solutions that may later serve in robotics to learn behaviorally diverse policies \cite{cully2017quality}. From a global perspective, the so-called \textit{discrimination-conscious by-design} paradigm collective refers to methodologies where discrimination detection and prevention is considered from the beginning of the design of the AI-based system through fairness \cite{hajian2016algorithmic}. Methods adopting this paradigm include discrimination-aware data mining \cite{pedreshi2008discrimination}, compositional fairness, interpretation of sanity checks and ablation studies.

In summary, diversity must be enforced both in the data from which models are learned and among the stakeholders, i.e., fostering the inclusion of minorities (practitioners and users) \cite{pisoni2021human,pisoni2023responsible} or using methodologies such as participatory design for accessibility \cite{diaz2020accessible}. Universal Design principles, which consider accessibility and ``design for all'' \cite{hleg2019ethics} during development, governance, policy and decision making processes is one way to facilitate AI life cycles that take into account what is beneficial for everyone, accounting for different conditions and situations, and no discrimination.

To further enable universal design and stakeholder participation, often feedback -- even after deployment -- is sought for stakeholder participation and consultation. One way to achieve this is through active learning for machine learning systems. Active learning allows for the integration of users' feedback while models are learned, and enables interactivity with the user, one of the goals targeted by human-centered AI \cite{shneiderman2022human} and AI for social good \cite{tomavsev2020ai}.

\subsubsection{Fairness}\label{sec:fairness}

The second dimension of this requirement is fairness, namely, techniques aimed to reduce the presence of unfair outputs elicited by AI-based systems. An unfair algorithm can be defined as that producing decisions that favor a particular group of people. Following the comprehensive view on this topic published in \cite{mehrabi2021survey}, biases leading to such unfair decisions can be propagated from the data to the AI algorithm (including measurement, omitted variable sampling, or representation biases, among others); from the algorithm to the user (as in algorithmic, popularity or evaluation biases); or from the user to the data (respectively, biases induced in the production of content, temporal, historical and/or social biases).

Fairness guarantees in the decisions of AI-based systems has been approached extensively in the literature, reporting bias targeting methods that can be classified in three large groups:
\begin{itemize}[leftmargin=*]
\item \emph{Pre-processing methods}, where the available data are transformed for the source of bias to be reduced and at best, removed.

\item \emph{In-processing methods}, which modify the learning algorithm of the model at hand (by e.g. changing the objective function at hand or imposing constraints to the optimization problem) so as to minimize the effect of biases in the training process. 

\item \emph{Post-processing methods}, which operate on the output of the model (for instance, by reassigning the predicted class for a query instance) without modifying its learning algorithm or the training data from which it was learned. 
\end{itemize}

In general, it is widely acknowledged that fairness can be achieved by sacrificing accuracy to a certain extent \cite{gu2022privacy}. However, it is also possible to debias machine learning models from the influence of spurious features to even improve their performance \cite{du2022towards}. Another trade-off is between fairness and privacy. Here, adversarial learning \cite{zhang2018mitigating} can simultaneously learn a predictor and an adversary that models a protected variable, and by minimizing the adversary capacity to predict this protected variable, accurate predictions can show less stereotyping of the protected variable, almost achieving equality of odds as a fairness notion.

An important concept to be acquainted with when dealing with fairness in AI-based systems is \textit{fairwashing}: as a risk of rationalization, fairwashing is the promotion of a false perception that a machine learning model respects ethical values through an outcome explanation and fairness metric \cite{aivodji2019fairwashing}. This makes it critical to characterize the manipulability of fairwashing \cite{aivodji2021characterizing}, as well as \textit{LaundryML} approaches \cite{aivodji2019fairwashing} to better audit unfair opaque models.

\subsubsection{Avoidance of unfair bias} \label{sec:bias}

Data and models can be exposed to a large set of potential bias-inducing phenomena. Ensuring diversity, representativeness and completeness in data and models needs to be a core part of the full AI life cycle (design, development and deployment phases of AI-based systems). Bias can be uncovered through proxy discrimination by models, since proxy variables are likely to be picked up, showing features as proxy that otherwise would not have been considered, i.e., zip codes in predictive policing \cite{o2017weapons}. As has been shown in the previous dimension, bias is not only algorithmic, but extends beyond the limits of models in a vicious cycle: starting with human activity bias, data bias, leads to sampling bias on the web (specially to be considered in the use of data to learn generative models), algorithmic bias, interaction bias and finally, self-selection bias that can revert back into the algorithm a second-order bias \cite{baeza2018bias}.

Bias mitigation techniques include several approaches \cite{parikh2019addressing,balayn2021managing}, from generic requirements and toolboxes \cite{silberg2019notes} to concrete taxonomies of bias \cite{playbook2020mitigating,gulati2022biased} at different stages of the AI life cycle \cite{suresh2021framework}. Different notions of fairness can be also defined \cite{mehrabi2021survey,barocas2019fairness}, including causal fairness -- which relies on causal relations and requires establishing causal graphs -- or counterfactual fairness. Causality can help debugging algorithmic bias mitigation or explaining models \cite{pearl66book}, e.g., causal mediation analysis can help uncover disparate impact of models by estimating the fairness associated to different explaining variables \cite{diaz2023gender}.

\subsection{Requirement 6: Societal and environmental wellbeing 
} \label{sec:p6}

\vspace{2mm}
\paragraph{WHAT does it mean?} AI-based systems should benefit all humankind, not only at the present time but also in future generations. Therefore, AI-based systems must be sustainable and environmentally friendly, so that the technological adoption of AI does not entail a progressive depletion of natural resources and maintains an ecological balance \cite{mariani2023trustworthy}. Therefore, dimensions supporting this requirement include sustainability and environmental friendliness, as well as a careful assessment of the social and societal impacts of AI.  

\paragraph{WHY is it important for trustworthiness?} AI systems should increase positive social change and enhance sustainability and ecological responsibility. Although they can be an effective tool to mitigate climate change \cite{rolnick2022tackling,salcedo2022analysis}, greenhouse gases emitted by the computationally intensive training processes of complex AI-based systems can exacerbate existing social and ethical challenges linked to AI \cite{cowls2021ai}. For instance, training only one single AI model can emit as many CO2 emissions as five cars in their entire lifetime. Computational and environmental costs grow proportionally to the complexity of the model in terms of its number of parameters \cite{hao2019training}. In particular, this study was done for large language models \cite{strubell2019energy} which cost about 8.4 tons per year, where an average carbon footprint for a person yearly is around 4. Although emissions are amortized over the model lifetime, the recent ChatGPT model was estimated to consume 1,287 MWh that translates into a cost of 522 tCO2e \cite{patterson2022carbon}. Therefore, energy and policy considerations are to be taken into account by institutions and companies implementing AI \cite{strubell2019energy}.

\paragraph{HOW can this requirement be met in practice?} This requirement is currently approached from two different angles: sustainability and environmental wellbeing (Subsection \ref{sec:sust}) and societal wellbeing (Subsection \ref{sec:societal}).

\subsubsection{Sustainability and environmental wellbeing} \label{sec:sust}

Sustainable AI \cite{wu2022sustainable} considers a holistic perspective that spans from models to data algorithms and hardware, and how software-hardware co-design can help mitigate carbon footprints of AI model life cycles (design, training and deployment stages). As mentioned previously, sustainable AI finds its motivation in the costly energy consumption of large AI models. Thus, sharing key learned lessons, best design practices, metrics, and standards is key for a sustainable development of AI systems. Technical contributions aimed to implement this requirement for the sustainability of AI are at the core of the \emph{Green AI} research area \cite{schwartz2020green}, which studies efficient and ecologically aware designs of AI-based algorithms, systems and assets.  

Many strategies to attain this requirement have been proposed over the years to reduce the environmental impact of AI models, with emphasis on those characterized by a large number of parameters and requiring long training latencies (e.g., deep neural networks). Among others: 
\begin{itemize}[leftmargin=*]
    \item Assessment of the environmental impact of AI-based systems with e.g., carbon footprint calculators\footnote{Greenhouse Gases Equivalencies Calculator, \url{https://www.epa.gov/energy/greenhouse-gases-equivalencies-calculator-calculations-and-references}, accessed on April 25th, 2023.} \cite{lacoste2019quantifying}. Evaluating the factors that influence AI’s greenhouse gas emissions is the first step towards mitigating its negative effects \cite{cowls2021ai}.
    \item Selection of the most relevant and necessary data, i.e., with smart data approaches \cite{maillo2020redundancy}.
    \item Model compression \cite{marino2023deep,mishra2020survey}, e.g. using quantization \cite{becking2022ecq}, distillation techniques \cite{hinton2015distilling,traore2019continual} or acceleration \cite{cheng2017survey} techniques.
    \item Consideration of efficiency as an evaluation metric and as a price tag to make models greener and more inclusive for researchers having limited resources \cite{schwartz2020green}. 
    \item Use of models that can rapidly adapt to new situations, domains and similar tasks by virtue of learning functionalities specifically devoted to this adaptation (e.g., multitask, few-shot learning, AutoML, meta-learning, neural architecture search or open-ended learning. This family of 
    GPAIS can provide more efficient, sustainable and less data depending AI systems. 
    \item Deployment of models on cloud computing servers fed with renewable energy sources, to minimize CO2 emissions. 
\end{itemize}

\subsubsection{Societal wellbeing} \label{sec:societal}

At the societal level, AI can improve social welfare. AI-based systems can perform routine tasks in an autonomous safer, and more efficient fashion, enhancing productivity and improving the quality of life of humankind. In the public administration AI can speed up processes, smooth administrative bottlenecks and save paperwork. Furthermore, it can aid policy making and help city planners, e.g., by visualizing the consequences of climate change, predicting future floods, or identifying urban heat islands. Possibilities for the society at large to benefit from AI developments have exploded in recent years with the progressive digitization of almost all sectors of activity. Infrastructure planning, health and hunger, equality and inclusion, education, economic empowerment, security and justice are among those sectors where AI can unleash its full potential to foster use cases of societal impact. 

Bringing such benefits of AI into practice is, therefore, a matter of leveraging such amounts of available data in AI-based systems. Such AI-based systems address learning tasks that solve a problem of societal impact, such as the ones exemplified above. However, since decisions issued by the AI-based system affect human beings and are subject to social scrutiny, other requirements of trustworthy AI become of utmost relevance, including fairness, privacy, transparency or human oversight. Above all, the importance of AI ethics and regulation becomes paramount in societal wellbeing, since decisions issued in use cases arising in education, justice and security have to comply with fundamental human rights and the legal restrictions in force.    

\subsection{Requirement 7: Accountability 
} \label{sec:p7} 

\vspace{2mm}
\paragraph{WHAT does it mean?} 
This last requirement of trustworthy AI systems imposes the provision of mechanisms to ensure responsibility and accountability for the development, deployment, maintenance and-or use of AI systems and their outcomes. Auditability, which enables the assessment of algorithms, data and design processes, plays a key role in accountability, namely, the attribution of the results of the actions that were taken based on the outcome of the AI-based system. Accountability, therefore, implies the minimization of harm and reporting of negative impact, the communication of design trade-offs to the user, and the implementation of adequate and accessible redress strategies associated to AI-based systems. Therefore, auditability and accountability are closely related to each other and lie at the core of \emph{responsible} AI systems, which are later discussed in Section \ref{sec:practice3}.

\paragraph{WHY is it important for trustworthiness?} 

The required auditability property of Trustworthy AI systems demands the development of practical tools \cite{zicari2022assess} that are capable of verifying desirable properties of neural networks such as stability, sensitivity, relevance or reachability \cite{ISO24029}, as well as metrics beyond explainability \cite{carvalho2019machine,hsiao2021roadmap,rosenfeld2021better,hoffman2018metrics,vitali2022survey}, such as on traceability, data quality and integrity. Auditability is becoming increasingly important when standards are being materialized touching upon all AI requirements. This includes IEEE, ISO/IEC and CEN/CENELEC, 
which are implementing concrete guidelines to apply trustworthy AI requirements in industrial setups (see \cite{mariani2023trustworthy,wahlster2020german} for an overview). At the national level, the German standardization road map on AI within DIN/DKE \cite{wahlster2020german} is a clear exponent of the standardization efforts made by different governments to dictate how practical AI-based systems should be audited. 

On the other hand, accountability is a key requirement to be able to recourse \cite{karimi2022towards} when an AI model contributes to making a proven wrong decision,  issuing explanations and recommendations to cases that are unfavorably treated by such decision. Accountability is a matter of compliance with ethical and legal standards, answerability, reporting and oversight, and attribution and enforcement of consequences \cite{novelli2023accountability}. Therefore, when framed under AI regulatory standards and ethical principles like the ones discussed in this work, accountability becomes crucial for AI-based systems to distribute cost, risks, burdens and liabilities among the different stakeholders participating in its life cycle.

\paragraph{HOW can this requirement be met in practice?} Similarly to other requirements, we next analyze how the different dimensions spanned by this requirement can be tackled in practice. In doing so, Subsection \ref{sec:acc} deals with accountability, whereas Subsection \ref{sec:audit} addresses auditability. The minimization and reporting of negative impacts is discussed in Subsection \ref{sec:minimisation}. Finally, Subsection \ref{sec:redress} describes methods for algorithmic redress.

\subsubsection{Accountability}\label{sec:acc} 

Mechanisms of accountability are especially relevant in high-risk scenarios, as they assign responsibility for decisions in the design, development and deployment phases of the AI system. Tools to attain this requirement involve algorithmic accountability policy toolkits (e.g., \cite{ainow2018aapt}), the post-hoc analysis of the output of the model (e.g. via local relevance attribution methods) or algorithms for causal inference and reasoning \cite{kim2021machine}. Since accountability is linked to the principle of fairness, it is closely related to risk management since unfair adverse effects can occur. Therefore, risks must be identified and mitigated transparently so they can be explained to and verified by third parties. Therefore, techniques and tools for auditing data, algorithms and design processes are required for accountable decisions issued by AI-based systems. An overview on 16 risk assessment frameworks is available in \cite{xia2023concrete}, whereas built-in derisking processes at design and development phases can be found in \cite{silberg2019notes,baquero2020derisking}. These processes operationalize risk management in machine learning pipelines, including explainability and bias mitigation. Another set of resources to tackle bias and fairness are discussed in \cite{silberg2019notes}.

Emerging trade-offs between requirements should be stated and assessed with regards to the risk they pose to ethical requirements and compromise of fundamental rights, since no AI system should be used when no risk-free trade-off for these can be found \cite{hleg2019ethics}. Consequently, AI models useful for accountability often involve multi-criteria decision making and pipelines at the \textit{MLOps} level that help delineate and inform such trade-offs to the user. 

\subsubsection{Auditability}\label{sec:audit} 

The AI Act has been interpreted as the European ecosystem to conduct AI auditing \cite{mokander2022conformity}. In the strict sense, the need for certifying systems that embed AI-based functionalities in their design is starting to permeate even within the international ISO standards for AI robustness. In such standards, formal methods for requirement verification or requirement satisfaction, typical of software engineering, are being extended towards verifying desirable properties of AI models. More specifically, in order to certify neural networks, properties such as stability, sensitivity, relevance or reachability are sought \cite{ISO24029}. 

In terms of auditing procedures, especially when the AI system interacts with users, grading schemes adapted to the use case \cite{holzinger2020measuring} are in need for validating models. Examples include the \emph{System Causability Scale} \cite{holzinger2020measuring} or the \emph{Muir Trust Scale} \cite{han2023communicating}, which are widely adopted in human robot interaction and robotics and rely on predictability (\textit{To what extent the robot behavior [the output of the AI-based system] can be predicted from moment to moment?}), reliability (\textit{To what extent can you count on the system to do its job?}), competence (\textit{What degree of faith does the user have on the system for it to cope with similar situations in the future?}) and trust (\textit{How much does the user trust the system overall?}). 

\subsubsection{Minimization and reporting of negative impacts and trade-offs}\label{sec:minimisation} 

The urgent need for developing stable and verifiable mechanisms for auditing AI-based systems becomes more relevant in the case of generative AI, which has grown so maturely that it is difficult to distinguish between human-created multimodal content and those generated by machines. If these are not properly identified, they can generate confusion and deception, which may have negative consequences for society, such as the manipulation of public opinion or the dissemination of fake news. 

A promising stream along these lines proposes to land the implementation of verifiable claims \cite{brundage2020toward}, which are defined as those falsifiable claims for which evidence and arguments can be provided to influence the probability that such claims are true. This proposal stems from the efforts of developers, regulators and other AI stakeholders, and the need to understand what properties of AI systems can be credibly demonstrated, through what means, and what trade-offs or commitments should and can be quantified. While the degree of certainty achievable varies across different claims and contexts, the idea is to demonstrate that greater degrees of evidence can be provided for claims about AI development than is typically done today to facilitate auditing them.

\subsubsection{Redress}\label{sec:redress} 

Lastly, once the risk has turned into a confirmed incident, it is paramount that the user is aware of the possibility to redress, preserving his/her trust when adverse or unfair impact takes place \cite{hleg2019ethics}. Redress is related to the concept of \textit{algorithmic recourse} \cite{karimi2022towards}, and consists of a procedure to correct or reverse an AI system outcome that is considered wrong. A key to trustworthy AI is ensuring adequate redress against decisions made by AI systems and by humans operating them through accessible mechanisms to their users when these fail, without forgetting vulnerable persons or collectives. Redress mechanisms are to be ensured, and complemented with accountability frameworks and disclaimers, since certification will obey particular application domains, and cannot replace responsibility. Machine unlearning \cite{bourtoule2021machine}, counterfactual explanations \cite{verma2020counterfactual} or the analysis of disparate impacts \cite{barocas2016big} can be also regarded as techniques that can support redress in AI-based systems.

\section{Trustworthy Artificial Intelligence from theory to practice and regulation: responsible Artificial Intelligence systems} \label{sec:applicability}

So far we have exposed the vision of trustworthy AI that has been tackled in most of the literature: from a theoretical point of view, and mainly based on principles and recommendations. In this section we highlight the importance of tackling trustworthy AI from a practical perspective. A clear mapping from trustworthy AI principles and requirements into operative protocols that can be automated, verified and audited does not always exist. To achieve this, the field needs blueprints and standard models to be adopted and standardized. In what follows we stress on the utmost importance of having practical regulatory scenarios (\emph{regulatory sandboxes}) and the final output of processes implementing HRAIs using trustworthy AI: a responsible AI system. 

According to this idea, the section is organized as follows. First, Subsection \ref{sec:practice3} defines the nuanced yet necessary notion of responsible AI systems, to comply with both trustworthy AI requirements and the law in force. Then, Subsection \ref{sec:practice1} describes the technical requirements that the implementation of HRAIs will legally require in practice. Then, Subsection \ref{sec:practice2} presents how these requirements are going to be evaluated by regulators and auditors through \emph{regulatory sandboxes}.  Subsection \ref{sec:practice4} examines whether all these steps can be connected and applied through a blueprint proposal to implement trustworthy AI in healthcare. \textcolor{black}{Finally, Subsection \ref{sec:65} examines the implications of new HRAIS and emerging AI systems, justifying the necessity of a dynamic regulation and flexible evaluation protocols to deal with new high-risk scenarios supported by these systems.}

\subsection{Responsible Artificial Intelligence systems}\label{sec:practice3}  

A little prior to \textit{trustworthy AI} is the term \textit{responsible AI}, which has been widely used quite as a synonym. However, it is necessary to make an explicit statement on the similarities and differences that can be established between trustworthy and responsible AI. The main aspects that make such concepts differ from each other is that responsible AI emphasizes the ethical use of an AI-based system, its auditability, accountability, and liability. 

In general, when referring to \emph{responsibility} over a certain task, the person in charge of the task assumes the consequences of his/her actions/decisions to undertake the task, whether they result to be eventually right or wrong. When translating this concept of responsibility to AI-based systems, decisions issued by the system in question must be accountable, legally compliant, and ethical. Other requirements for trustworthy AI reviewed in this manuscript (such as robustness or sustainability) are not relevant to responsibility. Therefore, trustworthy AI provides a broader umbrella that contains responsible AI and extends it towards considering other requirements that contribute to the generation of trust in the system. It is also worth mentioning that providing responsibility over AI products links to the provision of mechanisms for algorithmic auditing (\emph{auditability}), which is part of requirement 7 (\textit{Accountability}, Subsection \ref{sec:p7}). Stressing on the importance of a responsible development of AI, we now define the \emph{responsibility} associated to AI systems, following the discussed features.

\vspace{2mm}
\noindent\textbf{Definition}. A \textit{Responsible AI system} requires ensuring auditability and accountability during its design, development and use, according to specifications and the applicable regulation of the domain of practice in which the AI system is to be used. 

\vspace{2mm}
In the following we discuss in depth these features: 
\begin{enumerate}[leftmargin=*]
    \item \emph{Auditability}: As an element to aid accountability, a thorough auditing process aims to validate the conformity of the AI-based asset under target to 1) vertical or sectorial regulatory constraints; 2) horizontal or AI-wide regulations (e.g., EU AI Act); and 3) specifications and constraints imposed by the application for which it is designed. It is important to note that auditability refers to a property sought for the AI-based system, which may require transparency (e.g. explainability methods, traceability), measures to guarantee technical robustness, etc. This being said, the auditability of a responsible AI system may not necessarily cover all requirements for trustworthy AI, but rather those foretold by ethics, regulation, specifications and protocol testing {\color{black}adapted to the application sector (i.e., vertical regulation).}
    
    \item \emph{Accountability}: which establishes the liability of decisions derived from the AI system's output, once its compliance with the regulations, guidelines and specifications imposed by the application for which it is designed has been audited. Again, accountability may comprise different levels of compliance with the requirements for trustworthy AI defined previously.    
\end{enumerate}

In the context of the European approach and AI Act, this translates into a required pre-market use of regulatory sandboxes, and the adaptability of the requirements and regulation for trustworthy AI into a framework for the domain of practice of the AI system.

\subsection{Artificial Intelligence systems' compliance with regulation in high-risk scenarios} \label{sec:practice1}

It has been concluded in the previous section that the conformity of requirements are central for the definition of responsible AI systems. In Europe, regulatory requirements in force for the deployment of AI-based systems are prescribed based on the risk of such systems to cause harm. Indeed, the AI Act agreed by the European Parliament, the Council of the European Union, and the European Commission, is foreseen to set a landmark piece of legislation governing the use of AI in Europe and regulating this technology based on the definition of different levels of risks: minimal, limited and HRAIs. In these categories different requirements for trustworthy AI and levels of compliance are established, so that regulatory obligations are enforced therefrom. 

Furthermore, the European Commission has also asked the European Committee for Standardization (CEN), the European Committee for Electrotechnical Standardization (CENELEC) and the European Telecommunications Standards Institute (ETSI) to develop standards aimed to cover different practical aspects of AI systems, including foundational AI standards, data standards related to AI, Big Data and analytics, use cases and applications, governance implications of AI and computational approaches of AI. Ethical, societal concerns and AI trustworthiness also appear among the prioritized areas of these standardization bodies. 

Among these defined levels of risk associated to AI-based systems, those at the top of this classification (HRAIs) are subject to stringent obligations. HRAIs are demanded to comply with the AI Act through the fulfillment of the following seven requirements (AI Act, Chapter 2 \cite{AIA21}):
\begin{enumerate}[leftmargin=*]
\item \textit{Adequate risk assessment and mitigation systems} (Art. 9 - \textit{Risk management system}). 
\item \textit{High quality of the datasets feeding the system to minimize risks and discriminatory outcomes} (Art. 10 - \textit{Data and data governance}; Art. 9 - \textit{Risk management system}).
\item \textit{Logging of activity to ensure traceability of results} (Art. 12 - \textit{Record Keeping}; 20 - \textit{Automatically generated logs}).
\item \textit{Detailed documentation providing all information necessary on the system and its purpose for authorities to assess its compliance} (Art. 11 - \textit{Technical documentation};
Art. 12 - \textit{Record-keeping}).
\item \textit{Clear and adequate information to the user} (Art. 13 - \textit{Transparency}).
\item \textit{Appropriate human oversight measures to minimise risk} (Art. 14 - \textit{Human oversight}).
\item \textit{High level of robustness, security and accuracy} (Art. 15 - \textit{Accuracy, robustness and cybersecurity}).
\end{enumerate}

HRAIs must undergo conformity assessments before entering the EU market. One of the most complete guidance procedures assisting on complying with AI Act regulation is the \textit{CapAI} procedure for conducting conformity assessment of AI systems \cite{floridi2022capai}. It describes notions and metrics, checklists and other procedures to comply with the new legislation.

Since the AI Act imposes obligations on providers, importers, distributors, and users, the latter can be deemed as providers in certain cases. For instance, if a user slightly modifies or uses a ready-made AI-based product such as ChatGPT for a different purpose, this makes him/her become \textit{responsible} and accountable for the system's consequences, depending on the conditions that define HRAIs. This is why in order to realize trustworthy AI that is compliant with the law, we advocate for the development of \textit{responsible AI systems}, i.e., systems that not only make a responsible implementation that fulfills the requirements for trustworthy AI, but also comply with the AI regulation. 

In practice, HRAIs providers must work to make their assets meet these requirements, including post-market monitoring plans \cite{mokander2022conformity} (AI Act Art. 61 - \textit{Post-market monitoring by providers and post-market monitoring plan for high-risk AI systems}) to document the performance throughout the system's lifetime, in a way that vague concepts become verifiable criteria that strengthen the assessment safeguards and internal checks. Likewise, \textit{conformity assessments} (AI Act, Art. 19 and Art. 43) will be ruled by internationally harmonized testing principles, guaranteeing high-quality testing. These tests can depart from ad-hoc procedures and protocols for the domain at hand. This is the case of the German standardization roadmap on AI \cite{wahlster2020german}, which proposes conformity assessments based on several steps: calibration, inspection, audit, validation and verification. 

This need for harmonized testing protocols, monitoring plans and conformity assessment procedures is the main reason for the emergence of the concept of AI regulatory sandboxes, which are next detailed and discussed.

\subsection{Artificial Intelligence regulatory sandboxes: A challenge for auditing algorithms}
\label{sec:practice2}

Once requirements for HRAIs have been established, the remaining challenge is to make the AI system comply with them appropriately. Such requisites (AI Act, Chapter 2, Art. 8-15) motivate the need for a test environment where to audit AI-based systems by safe and harmonized procedures established by the latter. Regulatory sandboxes are indeed recommended by the AI Act (Chapter 5, Art. 53-54). Concretely, the AI Act establishes that algorithms should comply with regulation and can be tested in a safe environment prior to entering the market. This auditing process can be implemented via regulatory sandboxes.

\begin{figure*}
\centering
\includegraphics[width=2\columnwidth]{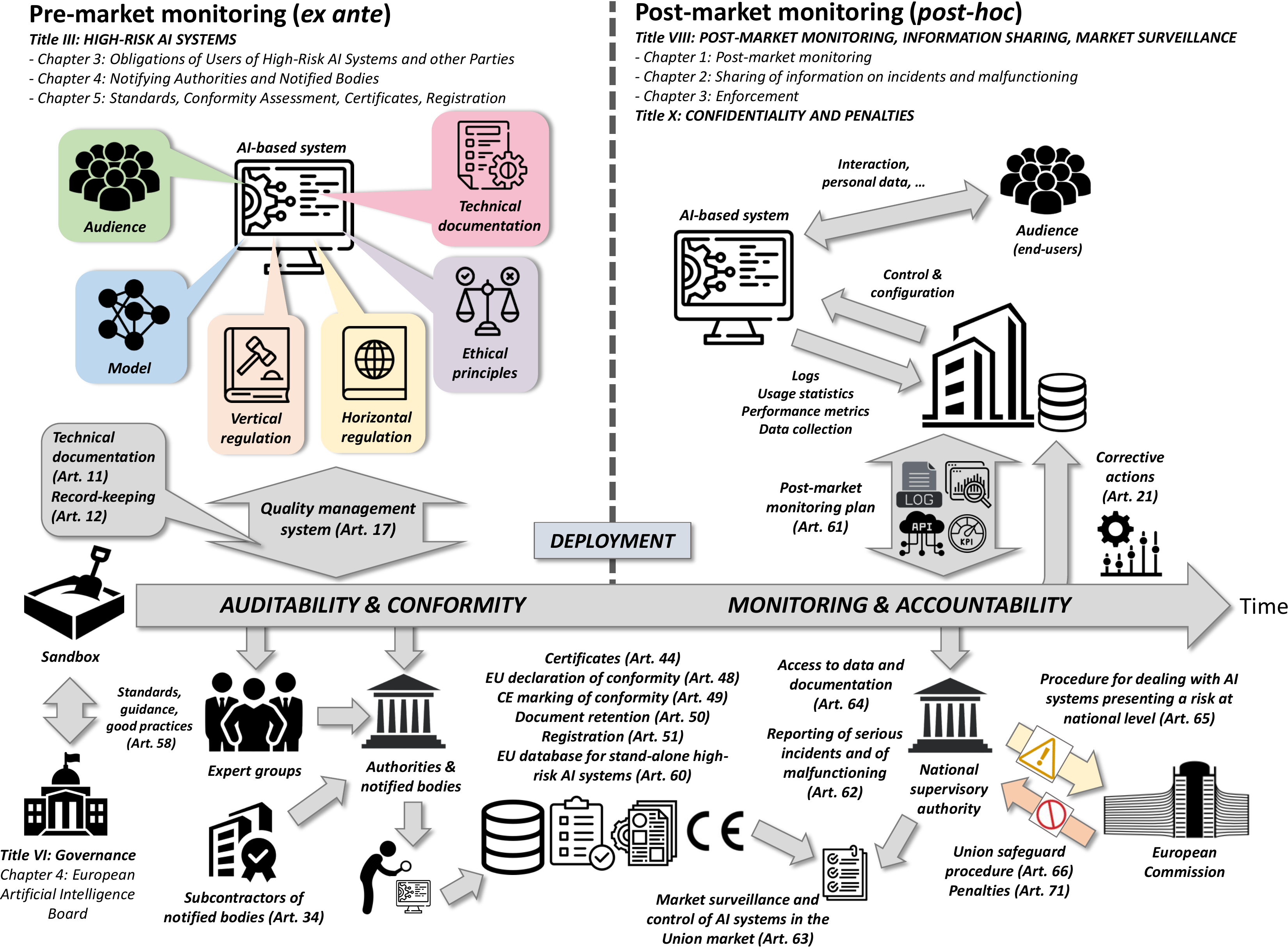}
\caption{{\color{black}Diagram showing the role of sandboxes before (\emph{ex-ante}) and after (\emph{post-hoc}) the AI-based system has been deployed in the market. Sandboxes permit to evaluate the conformity of the AI-based system w.r.t. technical specifications, horizontal \& vertical regulation, and ethical principles in a controlled and reliable testing environment. Once conformity has been verified, sandboxes can be used to interface with the deployed AI-based asset via the established monitoring plan, so that information about its post-market functioning can be collected and processed. This information is used by the national supervisory authority to evaluate the compliance: if needed, the authority asks for corrective actions and/or reports serious incidents/a continued lack of compliance to the European Commission. Articles in the AI Act related to each step are cited in the diagram.}}
\label{fig:sandboxfigPrePostMarket}
\end{figure*}

In order to successfully undertake AI auditing processes under the new regulation, industry, academia and governmental actors are forced to adapt their processes and teams to comply with the law. Regulatory sandboxes act as test beds and safe playgrounds that allow assessing the compliance of AI systems with respect to regulation, risk mitigation strategies, conformity assessments, accountability and auditing processes established by the law. {\color{black} Figure \ref{fig:sandboxfigPrePostMarket} illustrates the two stages where sandboxes play a crucial role: i) pre-market auditability and conformity check, and ii) post-market monitoring and accountability. The figure illustrates not only the different stakeholders participating in these two stages, but also the articles in the AI Act where each step within the process is described.}

In the current context of rapidly evolving AI products, sandboxes allow market stakeholders and business players to explore and experiment with new and innovative products, services or businesses under the supervision of a regulator. However, the idea of resorting to a sandbox to explore, evaluate and gauge complex technology is not new, nor exclusive of AI systems. They have already been used in other contexts to test and validate Fintech \cite{parenti2020regulatory} or Blockchain\footnote{Launch of the European Blockchain Regulatory Sandbox. \url{https://digital-strategy.ec.europa.eu/en/news/launch-european-blockchain-regulatory-sandbox}, accessed on April 25th, 2023.} technologies in the European Union. The objective of these controlled environments is to test innovative technologies for a limited time, \textcolor{black}{for innovators and regulators to cooperate\footnote{First regulatory sandbox on Artificial Intelligence presented: \url{https://digital-strategy.ec.europa.eu/en/news/first-regulatory-sandbox-artificial-intelligence-presented}}. The AI Act also contains measures with the aim to reduce the regulatory burden on Small and Medium Enterprises (SMEs) and startups, prioritize them, and to reduce their time to market by ensuring legislation can be implemented in two years. The intended goal is 
to support innovation and small-scale providers, getting apart from the \textit{regulation stifling innovation} critique.}

\textcolor{black}{The benefits of sandboxes is that they support the development, testing and validation of innovative AI systems under the direct supervision and guidance of competent authorities (AI Act Art. 53). Furthermore, they allow \emph{experimenting by derogation} (by putting aside certain rules or laws), and \emph{experimentation by devolution}, which requires broad supra/national frameworks to establish guidelines that empower and help local governments to establish a regulation in a particular area. This enables differences among government levels by considering local preferences and needs as a means to stimulate innovative policies.}

\textcolor{black}{When it comes to the challenges faced by sandboxes, there is a concern for the lack of proper methodological assessments that are indicative of the possible impact of AI on the society \cite{pop2021sandboxes}. This concern fosters the need for cross-border and multi-jurisdictional regulatory sandbox standardization \cite{yordanova2022eu}, as well as generic AI standardization \cite{josep2023ai}. Governments will have to find a balance between EU coordination and national procedures to avoid conflicts in the implementation of the regulation \cite{madiega2022artificial}. Specifically in the AI Act (Art. 53), participants in the sandbox remain liable under applicable liability legislation. Eligibility criteria and participants obligations and rights is to be set up in implementing acts. }

\textcolor{black}{Derived from the above challenge, we note that sandboxes are still far from maturity. This leads to two main aspects that remain unresolved:} 1) the design of sandboxes with guidelines that rapidly and effectively permit algorithmic auditing; and 2) the development of intelligent systems for high-risk scenarios that are validated through the necessary auditing processes. Important efforts are currently driven towards addressing these aspects as two additional fundamental challenges. At European level, Spain is leading a pilot to set up a regulatory sandbox according to the European AI Act legislation. 

\textcolor{black}{Together with sandboxes to work in practice, additional future mechanisms will include the certification or quality control within a regulatory framework. In this sense, Spain is starting to develop a national seal of quality to certify the security and quality of AI technology used in Spain. In cooperation with industry, they will set up the technical criteria for companies to obtain this seal, and develop tools to facilitate the certification process, e.g., developing self-assessment software. Several companies will be open the possibility to grant the seal, which will be voluntary for AI companies to obtain. At the international level, one effort towards this end is the IEEE \textit{CertifAIEd} program\footnote{IEEE \textit{CertifAIEd}: \url{https://engagestandards.ieee.org/ieeecertifaied.html}, accessed on June 6th, 2023.} to assess ethics of Autonomous Intelligent Systems via certification guidance, assessment and independent verification. This mark is meant for IEEE authorized assessors and certifiers to perform an independent review and verification to grant a mark and certificate based on ontological specifications for Ethical Privacy, Algorithmic Bias, Transparency, and Accountability.}

\textcolor{black}{We expect that the first experiences and results of running regulatory sandboxes and their alignment with certification activities will permit to learn lessons, to improve AI systems and eventually, to support the progressive proliferation of responsible AI systems deployed in practical scenarios. We believe that sandbox assessment should be periodically performed by independent and impartial assessment bodies to certificate and audit AI systems during their lifetime.}

\subsection{Practical case study in Artificial Intelligence for healthcare }\label{sec:practice4}

At the time of writing (April 2023), the AI Act regulation draft is constantly being updated through different amendments, due in part to new versions of AI products accessible to the general public. Concerned with how essential is the translation of principles and regulation into specific processes, it becomes necessary to have blueprint models and protocols that serve to assess how trustworthy AI systems are.

\begin{figure}[htb]
   \centering
   \includegraphics[width=0.95\linewidth]{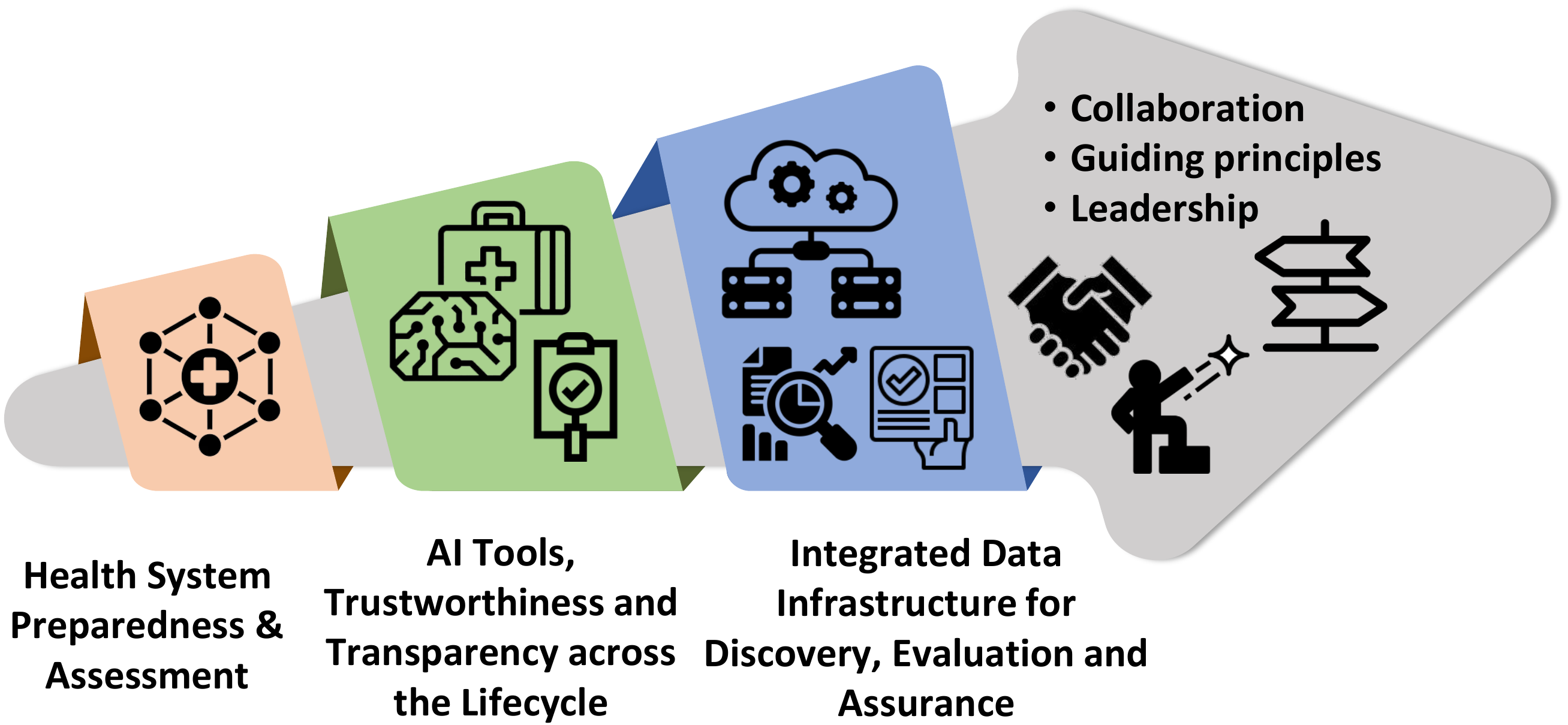}
 \caption{The Coalition for Health AI (CHAI) Blueprint for Trustworthy AI implementation guidance and assurance \cite{chai2023blueprint}. It considers obligations on reliability and testability, transparency, mitigation of biases particular to the domain, privacy, security and resilience, among other aspects.}
   \label{fig:CHAI_health_TAI_case}
 \end{figure}

The blueprint for \textit{Trustworthy AI Implementation Guidance and Assurance for Healthcare} is one step taken in this direction. Figure \ref{fig:CHAI_health_TAI_case} shows the proposal by the coalition for health AI \cite{chai2023blueprint}, based on collaboration, guiding principles and leadership actions. It is 
aligned with the \textit{AI risk management framework} from the National Institute of Standards and Technology (NIST). 

In particular, to ensure trustworthiness this framework describes four key functions \cite{chai2023blueprint}:

\begin{itemize}[leftmargin=*]
    \item \textit{Mapping} the framing AI risks, 
    \item \textit{Measuring} quantitatively and qualitatively these risks and impacts,
    \item \textit{Managing} the allocation of risk resources, and a cross-cutting, 
    \item \textit{Governance} via  risk management.
    
\end{itemize}

Based on these functions, they define values to set the \textit{key elements} of trustworthy AI in healthcare \cite{chai2023blueprint}: 
\begin{enumerate}[leftmargin=*]
    \item \emph{Useful: valid and reliable, testable, usable and beneficial}. These values are closely linked to social wellbeing (Requirement 6, Subsection \ref{sec:p6}) and auditability (Requirement 7, Subsection \ref{sec:p7}).
    \item \emph{Safe}, which is related to technical robustness and safety (Requirement 2, Subsection \ref{sec:p2}). 
    \item \emph{Accountable} and \emph{transparent}, with clear connections to accountability (Requirement 7, Subsection \ref{sec:p7}) and transparency (Requirement 4, Subsection \ref{sec:p4}).
    \item \emph{Explainable} and \emph{interpretable}, echoing several dimensions of the transparency requirement.
    \item \emph{Fair with harmful bias managed} (systemic bias, computational and statistical biases and human-cognitive biases). The consideration of fairness and the focus on the management of consequences of harmful biases are present in requirement 5 (Diversity, non-discrimination and fairness, Subsection \ref{sec:p5}), particularly in the avoidance of unfair biases. Furthermore, requirement 7 (accountability, Subsection \ref{sec:p7}) also regards the minimization of negative impacts, either due to harmful biases or to other consequences.
    \item \emph{Secure} and \emph{resilient}, which relate to the dimension of traceability (requirement 4, Subsection \ref{sec:p4}) and technical robustness (Requirement 2, Subsection \ref{sec:p2}).
    \item \emph{Privacy-enhanced}, which is coupled with requirement 3 -- Privacy and data governance (Subsection \ref{sec:p3}).
\end{enumerate}

It is important to underscore that some dimensions of the requirements for trustworthy AI discussed in Section \ref{sec:tai} are not reflected (at least, explicitly) in the above list of values, e.g. environmental wellbeing or reproducibility. This resounds with our formulated definition of a responsible AI system, showing that a system as such, depending on its domain of application, may require different degrees of compliance with the requirements for trustworthiness.

The blueprint analyzed in \cite{chai2023blueprint} recognizes the difficulties on building ecosystems when multiple guidelines are left out in the wild without a standardization consensus. It calls for mapping socio-technical scenarios to resolve tensions among principles, an ISO-based approach to \textit{professional responsibility}, and institutionalizing trustworthy AI Systems (that is, responsible AI systems).

As a follow-up of the blueprint, the coalition for health AI \cite{chai2023blueprint} suggests:  
\begin{itemize}[leftmargin=*]
    \item Setting up an engaged assurance lab and advisory service independent infrastructure.

    \item Institutionalizing trustworthy AI systems (responsible AI systems).  

    \item Promoting a \emph{Coalition of the Willing} through interesting strategies that can be applied in health AI to drive a positive change.
\end{itemize}

Other practical frameworks exist. They count with strategies to implement ethics and the governance of AI systems in health to separate the factors affecting trustworthy medical AI into design (data and algorithm aspects) and application. This is done through controls strategies \cite{zhang2023ethics} at both design and application phases. First, the ethical governance system departs from social needs and ethical values, which lead to ethical principles to be enforced at the research stage. After that, those principles guide the ethical norms that allow performing risk assessment, and later make the law and regulation concrete. 
In particular, the framework in \cite{amann2022explain} aims at avoiding situations that can have dire consequences for patients. For instance, integrating the applied ethics Z-Inspection{\textregistered} \cite{zicari2021z} process to map and assess tensions in socio-technical scenarios in trustworthy AI. \textcolor{black}{Another proposal \cite{muller2021ten} formulates ten commandments (i.e., high-level ethical principles) that should be met by medical AI-based systems. Such commandments are formulated without the theoretical aspects underneath for the sake of an easier understanding and verification of all stakeholders involved in this domain. European fundamental rights also provide legal and ethical guidelines for the adoption, development and application of medical AI \cite{stoger2021medical}.} These strategies and the blueprint are advancing the research in the area, and results will be extensible to other domains to attain trustworthy AI. 

Despite the clear establishment of the seven requirements for HRAIs within the AI Act described in Section \ref{sec:practice1}, the particular implementation steps to be taken within a particular area of application often remain under-specified. It becomes evident that the AI-based system is stringently dependent on the sector of application, as well as on the coupling of the requirements for trustworthy AI to justify existing regulations and standards. Therefore, for a given domain of practice, an overarching consideration of the complete scenario is needed from the Trustworthy AI practical point of view. The field needs to further specify legal requirements, risk assessment tools for the ethical impact of the AI system, data privacy and data governance models, ad-hoc risk management systems and conformity assessments, and rest of essential elements evaluated in the regulatory sandboxes testing the scenario. This may also spur the emergence of generic trustworthy AI frameworks (regulatory sandboxes) that can be potentially adapted to different domains, as it is discussed in \cite{baker2021taii}.

\subsection{\textcolor{black}{Urgent needs for emerging AI systems, dynamic regulation, and evaluation protocols}} \label{sec:65}

\textcolor{black}{The widespread use and repercussion of the achievements of emerging AI systems, such as GPAIS or neuroscience technology, have brought to the public arena the potentials and implications of new high-risk scenarios supported by these technological advances. In this section we discuss potential issues to be tackled to regulate new HRAIs as well as future emerging AI systems. We discuss and argue that regulation should be dynamic and malleable to establish the boundaries of new high-risk scenarios supported by technological AI advances. Likewise, we also highlight the need for flexible evaluation procedures that can be adapted in an agile way to cope with the fast evolution of AI systems.}

\textcolor{black}{Indeed, the rapid pace at which AI evolves over time can unexpectedly give rise to new high-risk scenarios beyond those defined by regulation, such as the AI Act (Section \ref{sec:practice1}). This requires regulatory protocols to cope with new emerging applications. In the case of the European AI Act, on 11th May 2023, MEPs endorsed new transparency and risk-management rules for AI systems\footnote{AI Act: a step closer to the first rules on Artificial Intelligence, \url{https://www.europarl.europa.eu/news/en/press-room/20230505IPR84904/ai-act-a-step-closer-to-the-first-rules-on-artificial-intelligence}, accessed on June 6th, 2023}. MEPs expanded the classification of high-risk areas to include those that could compromise or harm people’s health, safety, fundamental rights or the environment. Such revised classification of high-risk scenarios also considered AI systems used to influence voters in political campaigns, as well as recommender systems (with more than 45 million users) utilized by social media platforms. Intrusive and discriminatory uses of AI-based biometric systems have been also identified as prohibited AI systems, such as:}
\textcolor{black}{
\begin{itemize}[leftmargin=*]
\item ``Real-time'' remote biometric identification systems in publicly accessible spaces;
\item ``Post'' remote biometric identification systems, with the only exception of law enforcement for the prosecution of serious crimes and only after judicial authorization;
\item Biometric categorization systems using sensitive characteristics (e.g. gender, race, ethnicity, citizenship status, religion, political orientation);
\item Predictive policing systems (based on profiling, location or past criminal behavior);
\item Emotion recognition systems in law enforcement, border management, workplace, and educational institutions; and
\item Indiscriminate scraping of biometric data from social media or CCTV footage to create facial recognition databases (violating human rights and right to privacy).
\end{itemize}}

\textcolor{black}{
In this revision of the AI Act, MEPs have also proposed tailored regulatory regimes for new and fast-evolving developments in the field of AI and GPAIS. Since GPAIS are systems that have a wide range of possible uses without substantial modification and fine-tuning, generative foundation models are examples of rapidly evolving areas for which, if regulation is not set in place, consequences may be hard to revert. Such systems must guarantee robust the protection of fundamental rights, health and safety and the environment, democracy and rule of law. To this end, such emerging AI systems must assess and mitigate risks, comply with design, information and environmental requirements, and be registered in the EU database. Furthermore, additional transparency requirements have been demanded for generative foundation models such as GPT: they must inform that the content is generated by an AI model, the model must be designed to avoid generating illegal content and publishing summaries or copyrighted content used during training. Jurisdiction at national level will also need to be adapted to different considerations demanded by different sectors, e.g., the public sector or labor sector.}

\textcolor{black}{Another area in which regulation and technology yet have to advance is in copyright management of generated artwork produced by fundation models. Although the AI Act requires to disclose the use of copyrighted material in the training data, there is no current way to detect when AI generated content may be directly related to existing content protected by copyright, nor it is clear who owns the intellectual property of generative models outputs \cite{editorial2023writing,zirpoli2023generative}.}

\textcolor{black}{Besides GPAIS, other emerging AI-based technologies also require specialized adjustments of ongoing regulatory efforts. This is the case of neurotechnology, such as brain interfaces. The needs to handle novel applications never used before become evident by recent research \cite{tang2023semantic} that shows the potential of ``mind-reading'' \cite{reardon2023mind}. For instance, the study in \cite{tang2023semantic} shows the potential of leveraging language models as an autoregressive prior to generate novel sequences that can decode structured sequential information in the form of text from brain signals. Although the study of human imagination decoding shows human cooperation is required for the approach to work, this may not be a requisite in the future. Even if decoding is not accurate yet, these systems could be used maliciously.}

\textcolor{black}{These recent results attained by neurotechnology call for raising awareness about the risks posed by brain decoding technology, and for the design of regulation and policies to preserve fundamental rights such as mental privacy. A role model in this direction is the novel \textit{neurorights} regulation pioneered by Chile\footnote{NeuroRights Foundation, \url{https://neurorightsfoundation.org/, accessed on June 06th, 2023.}, which has taken a step towards the first Neurorights law in this country}. Their neuroprotection agenda (which is closely followed up by Spain) leads the way to the regulation of brain technology and legislation of advances in AI-supported medicine and science from a human rights' point of view. This is of utmost importance to avoid mental manipulation, as mental diseases are the grand pandemic of XXI century. If used appropriately, AI based technology applied to the diagnosis and treatment of mental diseases has a great potential to improve patients' quality of life.}

\textcolor{black}{In order for regulation to evolve synchronously with technological advances (and vice versa), evaluation procedures must be flexible and dynamically adaptable to accommodate the implications of these advances over time. More comprehensive evaluation tools are required to be adopted by the AI community of practitioners and users if we aspire to synergistic solutions that can complement governmental efforts. In particular, ML and AI model evaluation is normally based on leader-board benchmarks that do not always reflect reality, and may have a detrimental effect when they are not faithful to reality. A more accountable evaluation should consider aggregated metrics. Improperly constructed benchmarks may, for instance, reflect unrealistic overestimation of the capabilities of a model when predicting over minority classes. This may lead to hazards that end up obfuscating the real benefits of AI.}

{\color{black}To avoid these issues, important guidelines for robust evaluation practices \cite{burnell2023rethink} include:
\begin{enumerate}[leftmargin=*]
\item Granular performance reporting protocols with breakdowns across the features that have demonstrated affecting performance.
\item Designing benchmarks to test capabilities and to significantly vary on important features of the problem space and labeling instances to allow for granular analyses (e.g. as the Holistic Evaluation of Language Models benchmark \cite{liang2022holistic}).
\item Record all results, successful or failing, partial or fully traced, in supplementary material or public repositories for each run and validation split separately (e.g., in medical AI \cite{hernandez2020minimar}).
\item Enable researchers follow-up instance-level analyses by including data labels and annotations of those instances. 
\end{enumerate}

However, prospective evaluation methodologies as the one described above should be versatile and extensible to embrace and incorporate new performance metrics, evaluation protocols or even modeling tasks proposed along the years. But most importantly: their sought flexibility should not give rise to exceptional cases that would undermine the validity and applicability of regulations in force.}

\textcolor{black}{We conclude that given the fast pace at which AI is progressing in the last months, it is of paramount importance to have a dynamic regulation from a double perspective: the appearance of risk-based scenarios and the emergence of novel AI systems. Only in this way the regulation will facilitate the realization of responsible AI systems, in parallel to the development of methodologies for algorithmic auditing and the clearance of responsibilities in the use of such systems.}

\section{\textcolor{black}{From the Artificial Intelligence moratorium letter to regulation as the key for consensus}} \label{sec:debate}


{\color{black}At the time of writing, a global debate is held around the moratorium letter published by several renowned researchers calling for a pause in large AI experimentation\footnote{Future of Life Institute, Pause giant AI experiments: An open letter, \url{https://futureoflife.org/open-letter/pause-giant-ai-experiments/}, accessed on April 25th, 2023}. The letter can be interpreted as a contribution to pointing out the gap between the fast advance of high-powered AI systems and the regulation. The letter also highlights that:
\begin{quote}
{\color{black}\emph{``AI research and development should be refocused on making today's powerful, state-of-the-art systems more accurate, safe, interpretable, transparent, robust, aligned, trustworthy, and loyal.''}}
\end{quote}
}

{\color{black}Following up this moratorium letter, several declarations and written statements by reputed experts have been published to approach the AI conundrum between ethics, regulation and technological progress from different perspectives. Among them, we highlight the interview with G. Hinton\footnote{Geoffrey Hinton: "We need to find a way to control artificial intelligence before it’s too late", \url{https://english.elpais.com/science-tech/2023-05-12/geoffrey-hinton-we-need-to-find-a-way-to-control-artificial-intelligence-before-its-too-late.html}, accessed on June 4th, 2023}, in which he states that \emph{``We need to find a way to control artificial intelligence before it’s too late''}. Interestingly under the scope of this work, he has also underscored the different nature of intelligent systems when compared to human intelligence, and thereby the need for establishing regulation for these artificial systems:}
\begin{quote}
{\color{black}\emph{``Our brains are the result of evolution and have a series of integrated goals –- such as not hurting the body, hence the notion of damage; eating enough, hence the notion of hunger. Making as many copies of ourselves as possible, hence the sexual desire. Synthetic intelligence, on the other hand, hasn’t evolved: we’ve built it. Therefore, it doesn’t necessarily come with innate goals. So, the big question is, can we make sure that AI has goals that benefit us? This is the so-called alignment problem. And we have several reasons to be very concerned.''}}
\end{quote}

{\color{black}A similar line of thinking has been expressed by Harari\footnote{Yuval Noah Harari argues that AI has hacked the operating system of human civilisation, \url{https://www.economist.com/by-invitation/2023/04/28/yuval-noah-harari-argues-that-ai-has-hacked-the-operating-system-of-human-civilisation}, accessed on June 4th, 2023.}, emphasizing on the pressing immediacy at which regulation is needed to match the speed of AI technological development with the public use of AI systems:}
\begin{quote}
{\color{black}\emph{``We can still regulate the new AI tools, but we must act quickly. Whereas nukes cannot invent more powerful nukes, AI can make exponentially more powerful AI. The first crucial step is to demand rigorous safety checks before powerful AI tools are released into the public domain. Just as a pharmaceutical company cannot release new drugs before testing both their short-term and long-term side-effects, so tech companies shouldn’t release new AI tools before they are made safe. We need an equivalent of the Food and Drug Administration for new technology, and we need it yesterday.''}}
\end{quote}

{\color{black}Another example is the proposal made by Sam Altman (OpenAI co-founder) before the US Senate to defend the benefits of this revolutionary technology, claiming that AI regulation should ensure that the public has access to its many advantages\footnote{Written Testimony of Sam Altman Chief Executive Officer OpenAI, \url{https://www.washingtonpost.com/documents/0668f6f4-d957-4b94-a745-2aa9617d1d60.pdf?itid=lk_inline_manual_18}, accessed on June 4th, 2023.}}:
\begin{quote}
{\color{black}\emph{``[...] we are not alone in developing this technology. It will be important for policymakers to consider how to implement licensing regulations on a global scale and ensure international cooperation on AI safety, including examining potential intergovernmental oversight mechanisms and standard-setting.''}}
\end{quote}

{\color{black}Finally, the manifesto on AI risk supported by multiple scientists and notable figures in the AI landscape has claimed to elevate the mitigation of AI risks to the priority levels of other humanity-threatening fatalities\footnote{Statement on AI Risk, \url{https://www.safe.ai/statement-on-ai-risk}, accessed on June 04th, 2023.}:} 
\begin{quote}
{\color{black}\emph{``Mitigating the risk of extinction from AI should be a global priority alongside other societal-scale risks such as pandemics and nuclear war.''}}
\end{quote}


{\color{black}The controversy held around these and other statements is whether humanity is close to or far from the moment at which AI can pose a realistic threat to its own existence. Unquestionably, triggering this debate has ignited even further the need for ethical and regulatory frameworks that regulate whether and how AI-based systems can be trusted and used in practical setups. }

{\color{black} After this latter manifesto, the {\it{Center for AI Safety}} has just published a new document entitled {\it{Existing policy proposals aimed at present and future harms'}\footnote{Existing Policy Proposals Targeting Present and Future Harms, \url{https://https://www.safe.ai/post/three-policy-proposals-for-ai-safety}, accessed on June 07th, 2023.}}. The aim of this one-page document is to describe three proposals that, in our view, promote AI safety. We follow with a short analysis of these proposals:  
\begin{itemize}[leftmargin=*]
\item \emph{Legal Liability for AI harms}: The first issue highlighted in the document is the need for establishing improved legal liability frameworks for the accountability of damages caused by the application of AI systems. GPAIS are also referred in the document for the ill-advised implementation of legal exemptions to absolve GPAIS developers of liability, as such exemptions could unfairly shift the entire burden of responsibility from large corporations to smaller actors, users and communities lacking the necessary resources, access, and capabilities to effectively address and alleviate all risks.

\item \emph{Increased regulatory scrutiny}: The second problem emphasized in this document is the need for a greater regulatory inspection during the development of AI systems, extending beyond the application layer to encompass the entire product lifecycle. It underscores the importance of holding companies responsible for the data and design choices they make when developing these models. In line with this proposal, increased transparency and regulations over training data are crucial to address algorithmic bias effectively, and to prevent companies from unfairly leveraging copyrighted materials through data modeling without compensating their creators.

\item \emph{Human supervision of automated systems}: The third theme in the document is the importance of human oversight in the implementation of HRAIs. Human oversight can contribute to lessening potential concerns with bias and the propagation of false or misleading information through AI systems. An explicit reference is done to the EU's regulatory proposal, with a positive emphasis on the importance therein granted to the human oversight in the deployment of HRAIs.
\end{itemize}

Our position, as we put it in this manuscript, is that {\bf{"regulation is a key for consensus"}} among these diverging voices to cast light over the shadows of modern AI technologies. For this to occur, technologies, methodologies and tools supporting the development, auditability and accountability of responsible AI systems are of utmost importance to cope with high-risk scenarios and to meet regulatory constraints.

To finish this section, we pay attention to a final point made by the authors of the paper \cite{laux2022trustworthy}. Unfortunately, this road towards consensus is not exempt of their own risks. Indeed, conflating trust and trustworthiness with the acceptability of risks blurs the distinction between acceptability judgments made by domain experts and the trustworthiness of AI systems implemented in society \cite{laux2022trustworthy}. It has been argued that trust is improbable to be produced on demand and impossible on command, as \emph{``trust engineering''} may backfire and not achieve its goal. Focused on trust and trustworthiness in AI in the public sector, \cite{laux2022trustworthy} argues on the four acute challenges facing the European Commission’s attempt to signal the trustworthiness of AI through its proposed regulation: the uncertainty about the antecedents of perceived trust in public institutions that utilize AI; the threat of misalignment between trustworthiness and degrees of trust; concealed behavioral factors behind the acceptability of risks; and the need for impartial intermediaries.

Despite these and other curves in the road, regulation can be an unquestionable driving force to consolidate and put all these diverging voices on the same page. Regulation has favored consensus about the benefits and restrictions of technological advances that have evolved faster than expected, permeating quickly into the society (e.g., social networks, Internet or mobile communications). AI should not be an exception. There is still a long way to go before we have fully aligned AI technology and regulation, developing responsible AI systems adapted to each risk scenario and fully leveraging the latest advances in the field. For this to occur, the European regulatory model based on risk-based use case scenarios can serve as a guiding light for the maturity and implementation of ethical, legal and technical frameworks, fostering the creation of industrial and institutional instruments (e.g. AI sandboxes or AI ethics board \cite{schuett2023design}) that guarantee that AI-based products and services comply with their requirements.} 

\section{Concluding remarks} \label{sec:concl}

For years now, the ever-growing capabilities of AI-powered systems have stimulated debates about the impact, benefits, implications and risks brought by AI systems to the industry and society. The ground-breaking potential of large generative AI models such as ChatGPT and GPT4 has reinvigorated this debate, since their near general-purpose capabilities learned from multimodal data can support a wide variety of intended and unintended purposes and tasks, by generating content that is hardly distinguishable from that made by humans. This notorious advance has reinvigorated the relevance and momentum of trustworthy AI systems, particularly in what refers to 1) the ethical usage of these models, and 2) the need for regulatory directives that establish what, when and how AI systems can be adopted in practical applications.
 
In this context, this manuscript has shed light on the principles, pillars and requirements to be met by trustworthy AI systems to be considered as such. To this end, we have departed from mature regulation/supervisory frameworks developed around trustworthy AI (e.g. AI Act) to provide clear definitions of all related concepts, placing emphasis on what each requirement for trustworthiness in AI stands for, why they contribute to generating trust in the user of an AI-based system, and how such requirements can be met technically. Regarding the latter, a short tour over technological areas that can contribute to each of these requirements has been offered. Our study has also overviewed ethical principles for the development of AI, which establish an overarching set of recommendations that ensure that this discipline will be advanced under social and ethical standards. The study has been complemented by a discussion on practical aspects to be considered in the design, development and use of trustworthy AI systems, stressing on the importance of assessing their conformity to regulations (auditability) and explaining how their decisions are issued (accountability). These two practical aspects must be met by \emph{responsible} AI systems. 

Further along this line, accountability and explainability have permeated deeply into the recommendations recently issued for the development of trustworthy medical AI, a risk-critical sector in large demand for trust when embracing new technological advances. Our analysis of such recommendations has exposed that auditability and accountability are at the core of the guidelines proposed in this area; together with ethics, data governance and transparency. Medical AI exemplifies the paramount relevance of considering all these requirements for trustworthiness along the entire AI cycle.

\begin{tcolorbox}[breakable,notitle,boxrule=0pt,colback=gray!20,colframe=gray!20]
For a given domain of practice, we need to assess the complete scenario from the Trustworthy AI practical point of view, that is, all essential elements audited in regulatory sandboxes for scenario testing, together with clear accountability protocols. Above all, the development of responsible AI systems as the final output of the chain is essential and must be the goal for current AI designs and developments. 
\end{tcolorbox}

In summary, we hope that this paper serves as a reference for researchers, practitioners and neophytes who are new to the world of AI, with interest in trustworthy AI from a holistic perspective. A well-rounded analysis of what trust means in AI-based systems and its requirements as the one offered in this manuscript is a key for the design and development of responsible AI systems throughout their life cycle. We should not regulate scientific progress, but rather products and its usage. {\color{black}As we emphasize in this paper, regulation is the key for consensus, and for this purpose, trustworthy AI and responsible AI systems for high risk scenarios are imperative, as they will contribute to the convergence between technology and regulation}, the advance of science, the prosperity of our economies, and the good of humanity, subject to legal requirements and ethical principles.

\section{Acknowledgments}

N. D\'iaz-Rodr\'iguez is currently supported by a Marie Skłodowska-Curie Actions (MSCA) Postdoctoral Fellowship with agreement ID: 101059332 
and the Leonardo Scholarship for Researchers and Cultural Creators 2022 from the BBVA Foundation. J. Del Ser has received funding support from the Spanish \emph{Centro para el Desarrollo Tecnológico Industrial} (CDTI) through the AI4ES project, and from the Basque Government (\emph{Eusko Jaurlaritza}) through the Consolidated Research Group MATHMODE (IT1456-22). F. Herrera has received funding support from the Spanish \emph{Ministry of Science and Innovation} (grant PID2020-119478GB-I00). 

\section*{Declaration of competing interest}
The authors declare that they have no known competing financial interests or personal relationships that could have appeared to influence the work reported in this paper.

\bibliographystyle{elsarticle-num}
\bibliography{refs}

\begin{thebibliography}{100}
\expandafter\ifx\csname url\endcsname\relax
  \def\url#1{\texttt{#1}}\fi
\expandafter\ifx\csname urlprefix\endcsname\relax\def\urlprefix{URL }\fi
\expandafter\ifx\csname href\endcsname\relax
  \def\href#1#2{#2} \def\path#1{#1}\fi

\bibitem{ramesh2021zero}
A.~Ramesh, M.~Pavlov, G.~Goh, S.~Gray, C.~Voss, A.~Radford, M.~Chen,
  I.~Sutskever, Zero-shot text-to-image generation, in: International
  Conference on Machine Learning, PMLR, 2021, pp. 8821--8831.

\bibitem{saharia2022photorealistic}
C.~Saharia, W.~Chan, S.~Saxena, L.~Li, J.~Whang, E.~L. Denton, K.~Ghasemipour,
  R.~Gontijo~Lopes, B.~Karagol~Ayan, T.~Salimans, J.~Ho, D.~J. Fleet,
  M.~Norouzi,
  \href{https://proceedings.neurips.cc/paper_files/paper/2022/file/ec795aeadae0b7d230fa35cbaf04c041-Paper-Conference.pdf}{Photorealistic
  text-to-image diffusion models with deep language understanding}, in:
  S.~Koyejo, S.~Mohamed, A.~Agarwal, D.~Belgrave, K.~Cho, A.~Oh (Eds.),
  Advances in Neural Information Processing Systems, Vol.~35, Curran
  Associates, Inc., 2022, pp. 36479--36494.
\newline\urlprefix\url{https://proceedings.neurips.cc/paper_files/paper/2022/file/ec795aeadae0b7d230fa35cbaf04c041-Paper-Conference.pdf}

\bibitem{hleg2019ethics}
{European Commission High-Level Expert Group on AI}, {Ethics guidelines for
  trustworthy AI} (2019).

\bibitem{AIA21}
{European Union}, {Proposal for a Regulation of the European Parliament and of
  the Council Laying down harmonised rules on Artificial Intelligence
  (Artificial Intelligence Act) and amending certain Union Legislative Acts.
  COM/2021/206 final } (2021).

\bibitem{unesco2020Recommendation}
{UNESCO}, \href{en.unesco.org}{Recommendation on the ethics of artificial
  intelligence}, Digital Library UNESDOC (2020).
\newline\urlprefix\url{en.unesco.org}

\bibitem{benjamins2019responsible}
R.~Benjamins, A.~Barbado, D.~Sierra, {Responsible AI by design in practice},
  in: Proceedings of the Human-Centered AI: Trustworthiness of AI Models \&
  Data (HAI) track at AAAI Fall Symposium, 2019.

\bibitem{pisoni2021human}
G.~Pisoni, N.~D{\'\i}az-Rodr{\'\i}guez, H.~Gijlers, L.~Tonolli, Human-centered
  artificial intelligence for designing accessible cultural heritage, Applied
  Sciences 11~(2) (2021) 870.

\bibitem{stahl2018ethics}
B.~C. Stahl, D.~Wright, Ethics and privacy in {AI} and big data: Implementing
  responsible research and innovation, IEEE Security \& Privacy 16~(3) (2018)
  26--33.

\bibitem{coeckelbergh2020ai}
M.~Coeckelbergh, {AI ethics}, {MIT Press}, 2020.

\bibitem{coeckelbergh2020artificial}
M.~Coeckelbergh, Artificial intelligence, responsibility attribution, and a
  relational justification of explainability, Science and engineering ethics
  26~(4) (2020) 2051--2068.

\bibitem{wahlster2020german}
W.~Wahlster, C.~Winterhalter, German standardization roadmap on artificial
  intelligence, DIN/DKE, Berlin/Frankfurt (2020) 100.

\bibitem{edwards2022euAIA}
L.~Edwards, {The EU AI Act}: a summary of its significance and scope, Ada
  Lovelace Institute, Expert explainer Report (2022) 26.

\bibitem{campos2023definition}
S.~Campos, R.~Laurent, {A Definition of General-Purpose AI Systems: Mitigating
  Risks from the Most Generally Capable Models}, Available at SSRN 4423706
  (2023).

\bibitem{estevez2022glossary}
M.~Est{\'e}vez~Almenzar, D.~Fern{\'a}ndez~Llorca, E.~G{\'o}mez,
  F.~Martinez~Plumed, Glossary of human-centric artificial intelligence, Tech.
  Rep. JRC129614, Joint Research Centre (2022).

\bibitem{laux2022trustworthy}
J.~Laux, S.~Wachter, B.~Mittelstadt,
  \href{https://onlinelibrary.wiley.com/doi/abs/10.1111/rego.12512}{{Trustworthy
  artificial intelligence and the European Union AI act: On the conflation of
  trustworthiness and acceptability of risk}}, Regulation \& Governance
  n/a~(n/a).
\newblock \href
  {http://arxiv.org/abs/https://onlinelibrary.wiley.com/doi/pdf/10.1111/rego.12512}
  {\path{arXiv:https://onlinelibrary.wiley.com/doi/pdf/10.1111/rego.12512}},
  \href {https://doi.org/https://doi.org/10.1111/rego.12512}
  {\path{doi:https://doi.org/10.1111/rego.12512}}.
\newline\urlprefix\url{https://onlinelibrary.wiley.com/doi/abs/10.1111/rego.12512}

\bibitem{tjoa2020survey}
E.~Tjoa, C.~Guan, {A survey on explainable artificial intelligence (XAI):
  Toward medical XAI}, IEEE Transactions on Neural Networks and Learning
  Systems 32~(11) (2020) 4793--4813.

\bibitem{doran2017does}
D.~Doran, S.~Schulz, T.~R. Besold, {What does explainable AI really mean? A new
  conceptualization of perspectives}, arXiv preprint arXiv:1710.00794 (2017).

\bibitem{lipton2018mythos}
Z.~C. Lipton, The mythos of model interpretability: In machine learning, the
  concept of interpretability is both important and slippery, Queue 16~(3)
  (2018) 31--57.

\bibitem{hleg2020altai}
{European Commission High-Level Expert Group on AI}, {The Assessment List for
  Trustworthy Artificial Intelligence (ALTAI) for self assessment} (2020).

\bibitem{widmer2022towards}
C.~Widmer, M.~K. Sarker, S.~Nadella, J.~Fiechter, I.~Juvina, B.~Minnery,
  P.~Hitzler, J.~Schwartz, M.~Raymer, {Towards Human-Compatible XAI: Explaining
  Data Differentials with Concept Induction over Background Knowledge}, arXiv
  preprint arXiv:2209.13710 (2022).

\bibitem{lepri2021ethical}
B.~Lepri, N.~Oliver, A.~Pentland, Ethical machines: the human-centric use of
  artificial intelligence, Iscience (2021) 102249.

\bibitem{pisoni2023responsible}
G.~Pisoni, N.~D{\'\i}az-Rodr{\'\i}guez, Responsible and human centric
  {AI}-based insurance advisors, Information Processing \& Management 60~(3)
  (2023) 103273.

\bibitem{tomavsev2020ai}
N.~Toma{\v{s}}ev, J.~Cornebise, F.~Hutter, S.~Mohamed, A.~Picciariello,
  B.~Connelly, D.~C. Belgrave, D.~Ezer, F.~C. v.~d. Haert, F.~Mugisha, et~al.,
  Ai for social good: unlocking the opportunity for positive impact, Nature
  Communications 11~(1) (2020) 2468.

\bibitem{holzinger2016interactive}
A.~Holzinger, Interactive machine learning for health informatics: when do we
  need the human-in-the-loop?, Brain Informatics 3~(2) (2016) 119--131.

\bibitem{wef2019empowering}
{World Economic Forum}, Empowering {AI} leadership an oversight toolkit for
  boards of directors, Tech. rep. (2019).

\bibitem{wef2022empowering}
{World Economic Forum}, {Empowering AI Leadership: AI C-Suite Toolkit }, Tech.
  rep. (2022).

\bibitem{cambria2023survey}
E.~Cambria, L.~Malandri, F.~Mercorio, M.~Mezzanzanica, N.~Nobani, A survey on
  {XAI} and natural language explanations, Information Processing \& Management
  60~(1) (2023) 103111.

\bibitem{floridi2019establishing}
L.~Floridi, {Establishing the rules for building trustworthy AI}, Nature
  Machine Intelligence 1~(6) (2019) 261--262.

\bibitem{mariani2023trustworthy}
R.~Mariani, F.~Rossi, R.~Cucchiara, M.~Pavone, B.~Simkin, A.~Koene,
  J.~Papenbrock, {Trustworthy AI} -- {Part} 1, Computer 56~(2) (2023) 14--18.

\bibitem{chen2023ai}
P.-Y. Chen, P.~Das, {AI Maintenance: A Robustness Perspective}, Computer 56~(2)
  (2023) 48--56.

\bibitem{varshney2019trustworthy}
K.~R. Varshney, Trustworthy machine learning and artificial intelligence, XRDS:
  Crossroads, The ACM Magazine for Students 25~(3) (2019) 26--29.

\bibitem{yang2021generalized}
J.~Yang, K.~Zhou, Y.~Li, Z.~Liu, Generalized out-of-distribution detection: A
  survey, arXiv preprint arXiv:2110.11334 (2021).

\bibitem{ruospo2023survey}
A.~Ruospo, E.~Sanchez, L.~M. Luza, L.~Dilillo, M.~Traiola, A.~Bosio, A survey
  on deep learning resilience assessment methodologies, Computer 56~(2) (2023)
  57--66.

\bibitem{speakman2023detecting}
S.~Speakman, G.~A. Tadesse, C.~Cintas, W.~Ogallo, T.~Akumu, A.~Oshingbesan,
  Detecting systematic deviations in data and models, Computer 56~(2) (2023)
  82--92.

\bibitem{lesort2020continual}
T.~Lesort, V.~Lomonaco, A.~Stoian, D.~Maltoni, D.~Filliat,
  N.~D{\'\i}az-Rodr{\'\i}guez, Continual learning for robotics: Definition,
  framework, learning strategies, opportunities and challenges, Information
  fusion 58 (2020) 52--68.

\bibitem{abdar2021review}
M.~Abdar, F.~Pourpanah, S.~Hussain, D.~Rezazadegan, L.~Liu, M.~Ghavamzadeh,
  P.~Fieguth, X.~Cao, A.~Khosravi, U.~R. Acharya, et~al., A review of
  uncertainty quantification in deep learning: Techniques, applications and
  challenges, Information Fusion 76 (2021) 243--297.

\bibitem{parmar2023open}
J.~Parmar, S.~Chouhan, V.~Raychoudhury, S.~Rathore, Open-world machine
  learning: applications, challenges, and opportunities, ACM Computing Surveys
  55~(10) (2023) 1--37.

\bibitem{zimmermann2022increasing}
R.~S. Zimmermann, W.~Brendel, F.~Tramer, N.~Carlini,
  \href{https://openreview.net/forum?id=NkK4i91VWp}{Increasing confidence in
  adversarial robustness evaluations}, in: A.~H. Oh, A.~Agarwal, D.~Belgrave,
  K.~Cho (Eds.), Advances in Neural Information Processing Systems, 2022.
\newline\urlprefix\url{https://openreview.net/forum?id=NkK4i91VWp}

\bibitem{amodei2016concrete}
D.~Amodei, C.~Olah, J.~Steinhardt, P.~Christiano, J.~Schulman, D.~Man{\'e},
  {Concrete problems in AI safety}, arXiv preprint arXiv:1606.06565 (2016).

\bibitem{hendrycks2021unsolved}
D.~Hendrycks, N.~Carlini, J.~Schulman, J.~Steinhardt, Unsolved problems in ml
  safety, arXiv preprint arXiv:2109.13916 (2021).

\bibitem{mohseni2022taxonomy}
S.~Mohseni, H.~Wang, C.~Xiao, Z.~Yu, Z.~Wang, J.~Yadawa, Taxonomy of machine
  learning safety: A survey and primer, ACM Computing Surveys 55~(8) (2022)
  1--38.

\bibitem{gu2019badnets}
T.~Gu, K.~Liu, B.~Dolan-Gavitt, S.~Garg, Badnets: Evaluating backdooring
  attacks on deep neural networks, IEEE Access 7 (2019) 47230--47244.

\bibitem{hendrycks2021aligning}
D.~Hendrycks, C.~Burns, S.~Basart, A.~Critch, J.~Li, D.~Song, J.~Steinhardt,
  {Aligning AI with shared human values}, Proceedings of the International
  Conference on Learning Representations (ICLR) (2021).

\bibitem{o2017weapons}
C.~O'neil, Weapons of math destruction: How big data increases inequality and
  threatens democracy, Crown, 2017.

\bibitem{parikh2019addressing}
R.~B. Parikh, S.~Teeple, A.~S. Navathe, Addressing bias in artificial
  intelligence in health care, Jama 322~(24) (2019) 2377--2378.

\bibitem{bonawitz2019towards}
K.~Bonawitz, H.~Eichner, W.~Grieskamp, D.~Huba, A.~Ingerman, V.~Ivanov,
  C.~Kiddon, J.~Kone{\v{c}}n{\`y}, S.~Mazzocchi, B.~McMahan, et~al., Towards
  federated learning at scale: System design, Proceedings of Machine Learning
  and Systems 1 (2019) 374--388.

\bibitem{rodriguez2020federated}
N.~Rodr{\'\i}guez-Barroso, G.~Stipcich, D.~Jim{\'e}nez-L{\'o}pez, J.~A.
  Ruiz-Mill{\'a}n, E.~Mart{\'\i}nez-C{\'a}mara, G.~Gonz{\'a}lez-Seco, M.~V.
  Luz{\'o}n, M.~A. Veganzones, F.~Herrera, Federated learning and differential
  privacy: Software tools analysis, the {Sherpa.ai} {FL} framework and
  methodological guidelines for preserving data privacy, Information Fusion 64
  (2020) 270--292.

\bibitem{marcolla2022survey}
C.~Marcolla, V.~Sucasas, M.~Manzano, R.~Bassoli, F.~H. Fitzek, N.~Aaraj, Survey
  on fully homomorphic encryption, theory, and applications, Proceedings of the
  IEEE 110~(10) (2022) 1572--1609.

\bibitem{abadi2016deep}
M.~Abadi, A.~Chu, I.~Goodfellow, H.~B. McMahan, I.~Mironov, K.~Talwar,
  L.~Zhang, Deep learning with differential privacy, in: Proceedings of the
  2016 ACM SIGSAC Conference on Computer and Communications Security, 2016, pp.
  308--318.

\bibitem{universalguidelines}
{Public Voice coalition}, {Universal Guidelines for Artificial Intelligence},
  \url{https://thepublicvoice.org/ai-universal-guidelines/}, online [accessed
  April 20th, 2023] (2018).

\bibitem{ICO}
{Information Commissioner’s Office (ICO)}, {How to use AI and personal data
  appropriately and lawfully},
  \url{https://ico.org.uk/media/for-organisations/documents/4022261/how-to-use-ai-and-personal-data.pdf},
  online [accessed April 20th, 2023] (2022).

\bibitem{datagovernanceACT22}
E.~Union, {Regulation (EU) 2022/868 of the European Parliament and of the
  Council of 30 May 2022 on European data governance and amending Regulation
  (EU) 2018/1724 (Data Governance Act) } (2022).

\bibitem{dataACT22}
E.~Union, {Proposal for a REGULATION OF THE EUROPEAN PARLIAMENT AND OF THE
  COUNCIL on harmonised rules on fair access to and use of data (Data Act) }
  (2022).

\bibitem{arrieta2020explainable}
A.~{Barredo Arrieta}, N.~D{\'\i}az-Rodr{\'\i}guez, J.~Del~Ser, A.~Bennetot,
  S.~Tabik, A.~Barbado, S.~Garc{\'\i}a, S.~Gil-L{\'o}pez, D.~Molina,
  R.~Benjamins, et~al., {Explainable Artificial Intelligence ({XAI}): Concepts,
  taxonomies, opportunities and challenges toward responsible AI}, Information
  Fusion 58 (2020) 82--115.

\bibitem{haresamudram2023three}
K.~Haresamudram, S.~Larsson, F.~Heintz, Three levels of {AI} transparency,
  Computer 56~(2) (2023) 93--100.

\bibitem{perez2018systematic}
B.~P{\'e}rez, J.~Rubio, C.~S{\'a}enz-Ad{\'a}n, A systematic review of
  provenance systems, Knowledge and Information Systems 57 (2018) 495--543.

\bibitem{holzinger2021information}
A.~Holzinger, M.~Dehmer, F.~Emmert-Streib, R.~Cucchiara, I.~Augenstein,
  J.~Del~Ser, W.~Samek, I.~Jurisica, N.~D{\'\i}az-Rodr{\'\i}guez, Information
  fusion as an integrative cross-cutting enabler to achieve robust,
  explainable, and trustworthy medical artificial intelligence, Information
  Fusion 79 (2022) 263--278.

\bibitem{ALI2023101805}
S.~Ali, T.~Abuhmed, S.~El-Sappagh, K.~Muhammad, J.~M. Alonso-Moral,
  R.~Confalonieri, R.~Guidotti, J.~{Del Ser}, N.~Díaz-Rodríguez, F.~Herrera,
  {Explainable Artificial Intelligence (XAI): What we know and what is left to
  attain Trustworthy Artificial Intelligence}, Information Fusion (2023)
  101805.

\bibitem{ribeiro2016should}
M.~T. Ribeiro, S.~Singh, C.~Guestrin, {"Why should I trust you?" Explaining the
  predictions of any classifier}, in: Proceedings of the 22nd ACM SIGKDD
  International Conference on Knowledge Discovery and Data Mining, 2016, pp.
  1135--1144.

\bibitem{rajani2019explain}
N.~F. Rajani, B.~McCann, C.~Xiong, R.~Socher,
  \href{https://aclanthology.org/P19-1487}{Explain yourself! leveraging
  language models for commonsense reasoning}, in: Proceedings of the 57th
  Annual Meeting of the Association for Computational Linguistics, Association
  for Computational Linguistics, Florence, Italy, 2019, pp. 4932--4942.
\newblock \href {https://doi.org/10.18653/v1/P19-1487}
  {\path{doi:10.18653/v1/P19-1487}}.
\newline\urlprefix\url{https://aclanthology.org/P19-1487}

\bibitem{abhishek2022attribution}
K.~Abhishek, D.~Kamath, Attribution-based xai methods in computer vision: A
  review, arXiv preprint arXiv:2211.14736 (2022).

\bibitem{guidotti2019factual}
R.~Guidotti, A.~Monreale, F.~Giannotti, D.~Pedreschi, S.~Ruggieri, F.~Turini,
  Factual and counterfactual explanations for black box decision making, IEEE
  Intelligent Systems 34~(6) (2019) 14--23.

\bibitem{van2021evaluating}
J.~van~der Waa, E.~Nieuwburg, A.~Cremers, M.~Neerincx, Evaluating {XAI}: A
  comparison of rule-based and example-based explanations, Artificial
  Intelligence 291 (2021) 103404.

\bibitem{kaczmarek2022plenary}
K.~Kaczmarek-Majer, G.~Casalino, G.~Castellano, M.~Dominiak, O.~Hryniewicz,
  O.~Kami{\'n}ska, G.~Vessio, N.~D{\'\i}az-Rodr{\'\i}guez, Plenary: Explaining
  black-box models in natural language through fuzzy linguistic summaries,
  Information Sciences 614 (2022) 374--399.

\bibitem{bourgeais2022graphgonet}
V.~Bourgeais, F.~Zehraoui, B.~Hanczar, {GraphGONet: a self-explaining neural
  network encapsulating the Gene Ontology graph for phenotype prediction on
  gene expression}, Bioinformatics 38~(9) (2022) 2504--2511.

\bibitem{diaz2022explainable}
N.~D{\'\i}az-Rodr{\'\i}guez, A.~Lamas, J.~Sanchez, G.~Franchi, I.~Donadello,
  S.~Tabik, D.~Filliat, P.~Cruz, R.~Montes, F.~Herrera, {EXplainable
  Neural-Symbolic Learning (X-NeSyL) methodology to fuse deep learning
  representations with expert knowledge graphs: The MonuMAI cultural heritage
  use case}, Information Fusion 79 (2022) 58--83.

\bibitem{salewski2022clevr}
L.~Salewski, A.~Koepke, H.~Lensch, Z.~Akata, {CLEVR-X: A Visual Reasoning
  Dataset for Natural Language Explanations}, in: International Workshop on
  Extending Explainable AI Beyond Deep Models and Classifiers, Springer, 2022,
  pp. 69--88.

\bibitem{vilone2021notions}
G.~Vilone, L.~Longo, Notions of explainability and evaluation approaches for
  explainable artificial intelligence, Information Fusion 76 (2021) 89--106.

\bibitem{sevillano23}
I.~Sevillano-Garcia, J.~Luengo, F.~Herrera, {REVEL} framework to measure local
  linear explanations for black-box models: Deep learning image classification
  case study, International Journal of Intelligent Systems 2023 (2023) 8068569.

\bibitem{hupont2019demogpairs}
I.~Hupont, C.~Fern{\'a}ndez, Demogpairs: Quantifying the impact of demographic
  imbalance in deep face recognition, in: 14th IEEE International Conference on
  Automatic Face \& Gesture Recognition (FG 2019), IEEE, 2019, pp. 1--7.

\bibitem{fernando2021missing}
M.-P. Fernando, F.~C{\`e}sar, N.~David, H.-O. Jos{\'e}, Missing the missing
  values: The ugly duckling of fairness in machine learning, International
  Journal of Intelligent Systems 36~(7) (2021) 3217--3258.

\bibitem{gee2019explaining}
A.~H. Gee, D.~Garcia-Olano, J.~Ghosh, D.~Paydarfar, Explaining deep
  classification of time-series data with learned prototypes, in: CEUR workshop
  proceedings, Vol. 2429, NIH Public Access, 2019, p.~15.

\bibitem{cully2017quality}
A.~Cully, Y.~Demiris, Quality and diversity optimization: A unifying modular
  framework, IEEE Transactions on Evolutionary Computation 22~(2) (2017)
  245--259.

\bibitem{hajian2016algorithmic}
S.~Hajian, F.~Bonchi, C.~Castillo, Algorithmic bias: From discrimination
  discovery to fairness-aware data mining, in: Proceedings of the 22nd ACM
  SIGKDD International Conference on Knowledge Discovery and Data Mining, 2016,
  pp. 2125--2126.

\bibitem{pedreshi2008discrimination}
D.~Pedreshi, S.~Ruggieri, F.~Turini, Discrimination-aware data mining, in:
  Proceedings of the 14th ACM SIGKDD International Conference on Knowledge
  Discovery and Data Mining, 2008, pp. 560--568.

\bibitem{diaz2020accessible}
N.~D{\'\i}az-Rodr{\'\i}guez, G.~Pisoni, Accessible cultural heritage through
  explainable artificial intelligence, in: Adjunct Publication of the 28th ACM
  Conference on User Modeling, Adaptation and Personalization, 2020, pp.
  317--324.

\bibitem{shneiderman2022human}
B.~Shneiderman, Human-centered AI, Oxford University Press, 2022.

\bibitem{mehrabi2021survey}
N.~Mehrabi, F.~Morstatter, N.~Saxena, K.~Lerman, A.~Galstyan, A survey on bias
  and fairness in machine learning, ACM Computing Surveys (CSUR) 54~(6) (2021)
  1--35.

\bibitem{gu2022privacy}
X.~Gu, Z.~Tianqing, J.~Li, T.~Zhang, W.~Ren, K.-K.~R. Choo, Privacy, accuracy,
  and model fairness trade-offs in federated learning, Computers \& Security
  122 (2022) 102907.

\bibitem{du2022towards}
M.~Du, R.~Tang, W.~Fu, X.~Hu, Towards debiasing {DNN} models from spurious
  feature influence, in: Proceedings of the AAAI Conference on Artificial
  Intelligence, Vol.~36, 2022, pp. 9521--9528.

\bibitem{zhang2018mitigating}
B.~H. Zhang, B.~Lemoine, M.~Mitchell, Mitigating unwanted biases with
  adversarial learning, in: Proceedings of the 2018 AAAI/ACM Conference on AI,
  Ethics, and Society, 2018, pp. 335--340.

\bibitem{aivodji2019fairwashing}
U.~A{\"\i}vodji, H.~Arai, O.~Fortineau, S.~Gambs, S.~Hara, A.~Tapp,
  Fairwashing: the risk of rationalization, in: International Conference on
  Machine Learning, PMLR, 2019, pp. 161--170.

\bibitem{aivodji2021characterizing}
U.~A{\"\i}vodji, H.~Arai, S.~Gambs, S.~Hara, Characterizing the risk of
  fairwashing, Advances in Neural Information Processing Systems 34 (2021)
  14822--14834.

\bibitem{baeza2018bias}
R.~Baeza-Yates, Bias on the web, Communications of the ACM 61~(6) (2018)
  54--61.

\bibitem{balayn2021managing}
A.~Balayn, C.~Lofi, G.-J. Houben, Managing bias and unfairness in data for
  decision support: a survey of machine learning and data engineering
  approaches to identify and mitigate bias and unfairness within data
  management and analytics systems, The VLDB Journal 30~(5) (2021) 739--768.

\bibitem{silberg2019notes}
J.~Silberg, J.~Manyika, Notes from the {AI frontier}: Tackling bias in {AI}
  (and in humans), McKinsey Global Institute 1~(6) (2019).

\bibitem{playbook2020mitigating}
G.~Smith, I.~Rustagi,
  \href{https://haas.berkeley.edu/wp-content/uploads/UCB_Playbook_R10_V2_spreads2.pdf}{{Mitigating
  Bias in Artificial Intelligence, An Equity Fluent Leadership Playbook}},
  Berkeley Haas Center for Equity, Gender and Leadership (2020).
\newline\urlprefix\url{https://haas.berkeley.edu/wp-content/uploads/UCB_Playbook_R10_V2_spreads2.pdf}

\bibitem{gulati2022biased}
A.~Gulati, M.~A. Lozano, B.~Lepri, N.~Oliver, {BIASeD: Bringing Irrationality
  into Automated System Design}, in: Proceedings of the Thinking Fast and Slow
  and Other Cognitive Theories in AI (in AAAI 2022 Fall Symposium), Vol. 3332,
  2022.

\bibitem{suresh2021framework}
H.~Suresh, J.~Guttag, A framework for understanding sources of harm throughout
  the machine learning life cycle, in: Equity and access in algorithms,
  mechanisms, and optimization, 2021, pp. 1--9.

\bibitem{barocas2019fairness}
S.~Barocas, M.~Hardt, A.~Narayanan, Fairness and Machine Learning: Limitations
  and Opportunities, fairmlbook.org, 2019, \url{http://www.fairmlbook.org}.

\bibitem{pearl66book}
J.~Pearl, D.~Mackenzie, The Book of Why, Basic Books, 2018.

\bibitem{diaz2023gender}
N.~Díaz-Rodríguez, R.~Binkytė, W.~Bakkali, S.~Bookseller, P.~Tubaro,
  A.~Bacevičius, S.~Zhioua, R.~Chatila,
  \href{https://www.sciencedirect.com/science/article/pii/S0306457323000134}{Gender
  and sex bias in {COVID-19} epidemiological data through the lenses of
  causality}, Information Processing \& Management 60~(3) (2023) 103276.
\newblock \href {https://doi.org/https://doi.org/10.1016/j.ipm.2023.103276}
  {\path{doi:https://doi.org/10.1016/j.ipm.2023.103276}}.
\newline\urlprefix\url{https://www.sciencedirect.com/science/article/pii/S0306457323000134}

\bibitem{rolnick2022tackling}
D.~Rolnick, P.~L. Donti, L.~H. Kaack, K.~Kochanski, A.~Lacoste, K.~Sankaran,
  A.~S. Ross, N.~Milojevic-Dupont, N.~Jaques, A.~Waldman-Brown, et~al.,
  Tackling climate change with machine learning, ACM Computing Surveys (CSUR)
  55~(2) (2022) 1--96.

\bibitem{salcedo2022analysis}
S.~Salcedo-Sanz, J.~P{\'e}rez-Aracil, G.~Ascenso, J.~Del~Ser,
  D.~Casillas-P{\'e}rez, C.~Kadow, D.~Fister, D.~Barriopedro,
  R.~Garc{\'\i}a-Herrera, M.~Restelli, et~al., Analysis, characterization,
  prediction and attribution of extreme atmospheric events with machine
  learning: a review, arXiv preprint arXiv:2207.07580 (2022).

\bibitem{cowls2021ai}
J.~Cowls, A.~Tsamados, M.~Taddeo, L.~Floridi, The {AI} gambit: leveraging
  artificial intelligence to combat climate change -- opportunities,
  challenges, and recommendations, AI \& Society (2021) 1--25.

\bibitem{hao2019training}
K.~Hao, {Training a single AI model can emit as much carbon as five cars in
  their lifetimes}, MIT technology Review 75 (2019) 103.

\bibitem{strubell2019energy}
E.~Strubell, A.~Ganesh, A.~McCallum,
  \href{https://aclanthology.org/P19-1355}{Energy and policy considerations for
  deep learning in {NLP}}, in: Proceedings of the 57th Annual Meeting of the
  Association for Computational Linguistics, Association for Computational
  Linguistics, Florence, Italy, 2019, pp. 3645--3650.
\newblock \href {https://doi.org/10.18653/v1/P19-1355}
  {\path{doi:10.18653/v1/P19-1355}}.
\newline\urlprefix\url{https://aclanthology.org/P19-1355}

\bibitem{patterson2022carbon}
D.~Patterson, J.~Gonzalez, U.~H{\"o}lzle, Q.~Le, C.~Liang, L.-M. Munguia,
  D.~Rothchild, D.~R. So, M.~Texier, J.~Dean, The carbon footprint of machine
  learning training will plateau, then shrink, Computer 55~(7) (2022) 18--28.

\bibitem{wu2022sustainable}
C.-J. Wu, R.~Raghavendra, U.~Gupta, B.~Acun, N.~Ardalani, K.~Maeng, G.~Chang,
  F.~Aga, J.~Huang, C.~Bai, et~al., Sustainable {AI}: Environmental
  implications, challenges and opportunities, Proceedings of Machine Learning
  and Systems 4 (2022) 795--813.

\bibitem{schwartz2020green}
R.~Schwartz, J.~Dodge, N.~A. Smith, O.~Etzioni, {Green AI}, Communications of
  the ACM 63~(12) (2020) 54--63.

\bibitem{lacoste2019quantifying}
A.~Lacoste, A.~Luccioni, V.~Schmidt, T.~Dandres, Quantifying the carbon
  emissions of machine learning, arXiv preprint arXiv:1910.09700 (2019).

\bibitem{maillo2020redundancy}
J.~Maillo, I.~Triguero, F.~Herrera, Redundancy and complexity metrics for big
  data classification: Towards smart data, IEEE Access 8 (2020) 87918--87928.

\bibitem{marino2023deep}
G.~C. Marin{\'o}, A.~Petrini, D.~Malchiodi, M.~Frasca, Deep neural networks
  compression: A comparative survey and choice recommendations, Neurocomputing
  520 (2023) 152--170.

\bibitem{mishra2020survey}
R.~Mishra, H.~P. Gupta, T.~Dutta, A survey on deep neural network compression:
  Challenges, overview, and solutions, arXiv preprint arXiv:2010.03954 (2020).

\bibitem{becking2022ecq}
D.~Becking, M.~Dreyer, W.~Samek, K.~M{\"u}ller, S.~Lapuschkin, {ECQ:
  Explainability-Driven Quantization for Low-Bit and Sparse DNNs}, in:
  International Workshop on Extending Explainable AI Beyond Deep Models and
  Classifiers, Springer, 2022, pp. 271--296.

\bibitem{hinton2015distilling}
G.~Hinton, O.~Vinyals, J.~Dean, Distilling the knowledge in a neural network,
  arXiv preprint arXiv:1503.02531 (2015).

\bibitem{traore2019continual}
R.~Traor{\'e}, H.~Caselles-Dupr{\'e}, T.~Lesort, T.~Sun,
  N.~D{\'\i}az-Rodr{\'\i}guez, D.~Filliat, Continual reinforcement learning
  deployed in real-life using policy distillation and {Sim2Real} transfer, in:
  ICML Workshop on Multi-Task and Lifelong Reinforcement Learning, 2019.

\bibitem{cheng2017survey}
Y.~Cheng, D.~Wang, P.~Zhou, T.~Zhang, A survey of model compression and
  acceleration for deep neural networks, arXiv preprint arXiv:1710.09282
  (2017).

\bibitem{zicari2022assess}
R.~V. Zicari, J.~Amann, F.~Bruneault, M.~Coffee, B.~D{\"u}dder, E.~Hickman,
  A.~Gallucci, T.~K. Gilbert, T.~Hagendorff, I.~van Halem, et~al., How to
  assess trustworthy {AI} in practice, arXiv preprint arXiv:2206.09887 (2022).

\bibitem{ISO24029}
ISO/IEC, {ISO/IEC TR 24029-1, Information technology — Artificial
  Intelligence (AI) – Assessment of the robustness of neural networks - Part
  1: Overview}, \url{https://www.iso.org/standard/77609.html} (2021).

\bibitem{carvalho2019machine}
D.~V. Carvalho, E.~M. Pereira, J.~S. Cardoso, Machine learning
  interpretability: A survey on methods and metrics, Electronics 8~(8) (2019)
  832.

\bibitem{hsiao2021roadmap}
J.~H.-w. Hsiao, H.~H.~T. Ngai, L.~Qiu, Y.~Yang, C.~C. Cao, Roadmap of designing
  cognitive metrics for explainable artificial intelligence ({XAI}), arXiv
  preprint arXiv:2108.01737 (2021).

\bibitem{rosenfeld2021better}
A.~Rosenfeld, Better metrics for evaluating explainable artificial
  intelligence, in: Proceedings of the 20th International Conference on
  Autonomous Agents and MultiAgent Systems, 2021, pp. 45--50.

\bibitem{hoffman2018metrics}
R.~R. Hoffman, S.~T. Mueller, G.~Klein, J.~Litman, Metrics for explainable
  {AI}: Challenges and prospects, arXiv preprint arXiv:1812.04608 (2018).

\bibitem{vitali2022survey}
F.~Sovrano, S.~Sapienza, M.~Palmirani, F.~Vitali, A survey on methods and
  metrics for the assessment of explainability under the proposed {AI Act}, in:
  The Thirty-fourth Annual Conference on Legal Knowledge and Information
  Systems (JURIX), Vol. 346, IOS Press, 2022, p. 235.

\bibitem{karimi2022towards}
A.-H. Karimi, J.~von K{\"u}gelgen, B.~Sch{\"o}lkopf, I.~Valera, Towards causal
  algorithmic recourse, in: International Workshop on Extending Explainable AI
  Beyond Deep Models and Classifiers, Springer, 2022, pp. 139--166.

\bibitem{novelli2023accountability}
C.~Novelli, M.~Taddeo, L.~Floridi, Accountability in artificial intelligence:
  what it is and how it works, AI \& Society (2023) 1--12.

\bibitem{ainow2018aapt}
A.~Institute, \href{https://ainowinstitute.org/aap-toolkit.pdf}{{Algorithmic
  Accountability Policy Toolkit}} (2018).
\newline\urlprefix\url{https://ainowinstitute.org/aap-toolkit.pdf}

\bibitem{kim2021machine}
B.~Kim, F.~Doshi-Velez, Machine learning techniques for accountability, AI
  Magazine 42~(1) (2021) 47--52.

\bibitem{xia2023concrete}
B.~Xia, Q.~Lu, H.~Perera, L.~Zhu, Z.~Xing, Y.~Liu, J.~Whittle, Towards concrete
  and connected {AI} risk assessment ({C$^2$AIRA}): A systematic mapping study
  (2023).
\newblock \href {http://arxiv.org/abs/2301.11616} {\path{arXiv:2301.11616}}.

\bibitem{baquero2020derisking}
J.~A. Baquero, R.~Burkhardt, A.~Govindarajan, T.~Wallace, Derisking {AI} by
  design: How to build risk management into {AI} development, McKinsey \&
  Company (2020).

\bibitem{mokander2022conformity}
J.~M{\"o}kander, M.~Axente, F.~Casolari, L.~Floridi, Conformity assessments and
  post-market monitoring: A guide to the role of auditing in the proposed
  european {AI} regulation, Minds and Machines 32~(2) (2022) 241--268.

\bibitem{holzinger2020measuring}
A.~Holzinger, A.~Carrington, H.~M{\"u}ller, Measuring the quality of
  explanations: the system causability scale ({SCS}) comparing human and
  machine explanations, KI-K{\"u}nstliche Intelligenz 34~(2) (2020) 193--198.

\bibitem{han2023communicating}
Z.~Han, H.~Yanco, Communicating missing causal information to explain a
  robot’s past behavior, ACM Transactions on Human-Robot Interaction 12~(1)
  (2023) 1--45.

\bibitem{brundage2020toward}
M.~Brundage, S.~Avin, J.~Wang, H.~Belfield, G.~Krueger, G.~Hadfield, H.~Khlaaf,
  J.~Yang, H.~Toner, R.~Fong, et~al., Toward trustworthy {AI} development:
  mechanisms for supporting verifiable claims, arXiv preprint arXiv:2004.07213
  (2020).

\bibitem{bourtoule2021machine}
L.~Bourtoule, V.~Chandrasekaran, C.~A. Choquette-Choo, H.~Jia, A.~Travers,
  B.~Zhang, D.~Lie, N.~Papernot, Machine unlearning, in: IEEE Symposium on
  Security and Privacy (SP), IEEE, 2021, pp. 141--159.

\bibitem{verma2020counterfactual}
S.~Verma, V.~Boonsanong, M.~Hoang, K.~E. Hines, J.~P. Dickerson, C.~Shah,
  Counterfactual explanations and algorithmic recourses for machine learning: A
  review, in: NeurIPS 2020 Workshop: ML Retrospectives, Surveys \&
  Meta-Analyses (ML-RSA), 2020.

\bibitem{barocas2016big}
S.~Barocas, A.~D. Selbst, Big data's disparate impact, California Law Review
  (2016) 671--732.

\bibitem{floridi2022capai}
L.~Floridi, M.~Holweg, M.~Taddeo, J.~Amaya~Silva, J.~M{\"o}kander, Y.~Wen,
  {CapAI-A} procedure for conducting conformity assessment of {AI} systems in
  line with the {EU} artificial intelligence act, Available at SSRN 4064091
  (2022).

\bibitem{parenti2020regulatory}
R.~Parenti, Regulatory sandboxes and innovation hubs for fintech, Study for the
  Committee on Economic and Monetary Affairs, Policy Department for Economic,
  Scientific and Quality of Life Policies, European Parliament, Luxembourg
  (2020) 65.

\bibitem{pop2021sandboxes}
F.~Pop, L.~Adomavicius, Sandboxes for responsible artificial intelligence. eipa
  briefing september 2021. (2021).

\bibitem{yordanova2022eu}
K.~Yordanova, The {EU AI Act-Balancing} human rights and innovation through
  regulatory sandboxes and standardization (2022).

\bibitem{josep2023ai}
J.~{Soler Garrido}, S.~Tolan, I.~{Hupon Torres}, D.~{Fernandez Llorca},
  V.~Charisi, E.~{Gomez Gutierrez}, H.~Junklewitz, R.~Hamon, D.~{Fano Yela},
  C.~Panigutti, {AI Watch}: Artificial intelligence standardisation landscape
  update, Tech. rep., Joint Research Centre (Seville site) (2023).

\bibitem{madiega2022artificial}
T.~Madiega, A.~L. {Van De Pol},
  \href{https://www.europarl.europa.eu/RegData/etudes/BRIE/2022/733544/EPRS_BRI(2022)733544_EN.pdf}{{Artificial
  intelligence act and regulatory sandboxes. EPRS European Parliamentary
  Research Service. June 2022}} (2022).
\newline\urlprefix\url{https://www.europarl.europa.eu/RegData/etudes/BRIE/2022/733544/EPRS_BRI(2022)733544_EN.pdf}

\bibitem{chai2023blueprint}
{Coalition for Health AI (CHAI)},
  \href{https://www.coalitionforhealthai.org/papers/Blueprint\%20for\%20Trustworthy\%20AI.pdf}{Blueprint
  for trustworthy {AI} implementation guidance and assurance for healthcare}
  (2023).
\newline\urlprefix\url{https://www.coalitionforhealthai.org/papers/Blueprint\%20for\%20Trustworthy\%20AI.pdf}

\bibitem{zhang2023ethics}
J.~Zhang, Z.-M. Zhang, Ethics and governance of trustworthy medical artificial
  intelligence, BMC Medical Informatics and Decision Making 23~(1) (2023)
  1--15.

\bibitem{amann2022explain}
J.~Amann, D.~Vetter, S.~N. Blomberg, H.~C. Christensen, M.~Coffee, S.~Gerke,
  T.~K. Gilbert, T.~Hagendorff, S.~Holm, M.~Livne, et~al., To explain or not to
  explain?-- {Artificial} intelligence explainability in clinical decision
  support systems, PLOS Digital Health 1~(2) (2022) e0000016.

\bibitem{zicari2021z}
R.~V. Zicari, J.~Brodersen, J.~Brusseau, B.~D{\"u}dder, T.~Eichhorn, T.~Ivanov,
  G.~Kararigas, P.~Kringen, M.~McCullough, F.~M{\"o}slein, et~al.,
  Z-inspection{\textregistered}: a process to assess trustworthy {AI}, IEEE
  Transactions on Technology and Society 2~(2) (2021) 83--97.

\bibitem{muller2021ten}
H.~Muller, M.~T. Mayrhofer, E.-B. Van~Veen, A.~Holzinger, The ten commandments
  of ethical medical {AI}, Computer 54~(07) (2021) 119--123.

\bibitem{stoger2021medical}
K.~St{\"o}ger, D.~Schneeberger, A.~Holzinger, Medical artificial intelligence:
  the european legal perspective, Communications of the ACM 64~(11) (2021)
  34--36.

\bibitem{baker2021taii}
J.~Baker-Brunnbauer, {TAII Framework for Trustworthy AI systems}, ROBONOMICS:
  The Journal of the Automated Economy 2 (2021) 17.

\bibitem{editorial2023writing}
Editorials, Writing the rules in ai-assisted writing, Nature Machine
  Intelligence 469~(5) (2023) 469--469.
\newblock \href {https://doi.org/https://doi.org/10.1038/s42256-023-00678-6}
  {\path{doi:https://doi.org/10.1038/s42256-023-00678-6}}.

\bibitem{zirpoli2023generative}
C.~T. Zirpoli, Generative artificial intelligence and copyright law, United
  States Congressional Research Service, CRS Legal Sidebar, (February 23, 10922
  (5 pages) (2023).

\bibitem{tang2023semantic}
J.~Tang, A.~LeBel, S.~Jain, A.~G. Huth, Semantic reconstruction of continuous
  language from non-invasive brain recordings, Nature Neuroscience (2023) 1--9.

\bibitem{reardon2023mind}
S.~Reardon, Mind-reading machines are here: is it time to worry?, Nature
  617~(7960) (2023) 236--236.

\bibitem{burnell2023rethink}
R.~Burnell, W.~Schellaert, J.~Burden, T.~D. Ullman, F.~Martinez-Plumed, J.~B.
  Tenenbaum, D.~Rutar, L.~G. Cheke, J.~Sohl-Dickstein, M.~Mitchell, et~al.,
  Rethink reporting of evaluation results in ai, Science 380~(6641) (2023)
  136--138.

\bibitem{liang2022holistic}
P.~Liang, R.~Bommasani, T.~Lee, D.~Tsipras, D.~Soylu, M.~Yasunaga, Y.~Zhang,
  D.~Narayanan, Y.~Wu, A.~Kumar, et~al., Holistic evaluation of language
  models, arXiv preprint arXiv:2211.09110 (2022).

\bibitem{hernandez2020minimar}
T.~Hernandez-Boussard, S.~Bozkurt, J.~P. Ioannidis, N.~H. Shah, {MINIMAR
  (MINimum Information for Medical AI Reporting): developing reporting
  standards for artificial intelligence in health care}, Journal of the
  American Medical Informatics Association 27~(12) (2020) 2011--2015.

\bibitem{schuett2023design}
J.~Schuett, A.~Reuel, A.~Carlier, {How to design an AI ethics board}, arXiv
  preprint arXiv:2304.07249 (2023).

\end{thebibliography}
 
\appendix


\end{document}